\documentclass[journal]{IEEEtran}
\usepackage{epsfig,epsf}%,fullpage}
\usepackage{color}

\usepackage{bbold}
\usepackage{multirow}
\usepackage{latexsym}
\usepackage{amssymb}
\usepackage{amsmath}

\usepackage{amsfonts}
\usepackage[sort]{cite}
\usepackage{fullpage}
\usepackage{mathrsfs}
\usepackage{scalefnt}
\usepackage{extarrows}
\usepackage{setspace}
\usepackage{stfloats}
\usepackage{url}
\usepackage{hyphenat}
\usepackage{graphicx}
\usepackage[table]{xcolor}
\usepackage{bbm}
\usepackage{dsfont}
\usepackage{bm}
\usepackage{array}
\usepackage{wrapfig}
\usepackage[]{algpseudocode}
\usepackage[linesnumbered,ruled]{algorithm2e}

\usepackage{mathtools}
\allowdisplaybreaks[1]
\usepackage{epstopdf}
\newcommand\numberthis{\addtocounter{equation}{1}\tag{\theequation}}
\newsavebox\myboxA
\newsavebox\myboxB
\newlength\mylenA
\newcommand*\xoverline[2][0.75]{%
    \sbox{\myboxA}{$\m@th#2$}%
    \setbox\myboxB\null% Phantom box
    \ht\myboxB=\ht\myboxA%
    \dp\myboxB=\dp\myboxA%
    \wd\myboxB=#1\wd\myboxA% Scale phantom
    \sbox\myboxB{$\m@th\overline{\copy\myboxB}$}%  Overlined phantom
    \setlength\mylenA{\the\wd\myboxA}%   calc width diff
    \addtolength\mylenA{-\the\wd\myboxB}%
    \ifdim\wd\myboxB<\wd\myboxA%
       \rlap{\hskip 0.5\mylenA\usebox\myboxB}{\usebox\myboxA}%
    \else
        \hskip -0.5\mylenA\rlap{\usebox\myboxA}{\hskip 0.5\mylenA\usebox\myboxB}%
    \fi}
\makeatother  
\usepackage{enumerate}
\def\phi{\varphi}

\def\tendsto{\rightarrow}

\def\Tp{{\intercal}}

\newcommand{\comm}[1]{}
\def\bff{{\mathbf{f}}}
\def\bg{{\mathbf{g}}}

\def\bk{{\mathbf{k}}}

\def\bv{{\mathbf{v}}}

\def\bx{{\mathbf{x}}}
\def\by{{\mathbf{y}}}
\def\bz{{\mathbf{z}}}
\def\b0{{\mathbf{0}}}

\def\argmax{\mathop{\mathrm{argmax}}}

\def\bQ{{\mathbf{Q}}}

\def\cH{\mathcal{H}}
\def\cI{\mathcal{I}}

\def\cN{\mathcal{N}}

\def\cP{\mathcal{P}}

\def\cR{\mathcal{R}}

\def\cX{\mathcal{X}}
\def\cY{\mathcal{Y}}

\def \Re[#1]{\text{Re}\left(#1\right)}%{#1_R}
\def \Im[#1]{\text{Im}\left(#1\right)}%{#1_I}
\def \Cpx[#1]{\tilde{#1}}
\def \tx {x} % tx scalar case
\def \rx {y} % rx scalar case

\def \txv {\bx} % tx input vector
\def \rxv {\by} % rx output vector

\def \txm {X} % tx matrix
\def \rxm {Y} % rx matrix

\def \txA {\cX} % tx symbol alphabet
\def \rxA {\cY} % rx symbol alphabet
\def \chm {G} % channel matrix

\def \txvte[#1]{\txv_{\rm t,#1}} 
\def \rxvte[#1]{\rxv_{\rm t,#1}} 
\def \txvde[#1]{\txv_{\rm d,#1}} 
\def \rxvde[#1]{\rxv_{\rm d,#1}} 
\def \nvbe[#1]{\nv_{\rm b,#1}} 

\def \txde[#1]{\tx_{\rm d,#1}} 
\def \rxde[#1]{\rx_{\rm d,#1}} 

\def \txta[#1]{\tx_{{\rm t},#1}} 
\def \txda[#1]{\tx_{{\rm d},#1}} 

\def \txvta[#1]{\txv_{{\rm t},#1}} 
\def \rxvta[#1]{\rxv_{{\rm t},#1}} 
\def \txvda[#1]{\txv_{{\rm d},#1}} 
\def \rxvda[#1]{\rxv_{{\rm d},#1}} 
\def \nvba[#1]{\nv_{{\rm d},#1}} 
\def \rxta[#1]{\rx_{{\rm t},#1}} 
\def \txda[#1]{\tx_{{\rm d},#1}} 
\def \rxda[#1]{\rx_{{\rm d},#1}}

\def \ptrve[#1]{\hat{p}_{(\txvte[#1],\rxvte[#1])}}

\def \onev[#1]{\underline{1}_{#1}}% one vector
\def \MuI {I} % mutual information
\def \Ent {H} % entropy
\def \EntR {\cH}
\def \EntInn {{\cH{'}}}

\def \E {\mathbb{E}} % expectation
\def \Pr {{P}} % probability
\def \chv {\bg} % channel vector

\def \chs {g} % channel scalar
\def \tn {M} % tx antenna numbers
\def \rn {N} % rx antenna numbers
\def \ns  {v}  % noise scalar

\def \nv {\bv} % noise vector
\def \Tb {B} %total blocklength
\def \Tt {T} %block training number
\def \sign {\text{sign}}
\DeclarePairedDelimiter{\ceil}{\lceil}{\rceil}
\DeclarePairedDelimiter{\floor}{\lfloor}{\rfloor}

\def \ratio {\alpha}
\def \qha[#1]{q_{{ \chs},#1}}
\def \Qha[#1]{Q_{{\rm \chs},#1}}

\def \qxa[#1]{q_{{ x},#1}}
\def \Qxa[#1]{Q_{{\rm x},#1}}

\def \bQha[#1]{\bQ_{{\rm h},#1}}
\def \bQxa[#1]{\bQ_{{\rm x},#1}}

\def \Ropt {\cR_{\rm opt}}

\def \dy {\varepsilon}
\def \dx {\delta}

\def \iid {{\it iid}\xspace}

\def \rxdequa[#1]{\tilde{\rx}_{{\rm d},#1}}

\def \htxta[#1]{\hat{\tx}_{{\rm t},#1}} 
\def \htxda[#1]{\hat{\tx}_{{\rm d},#1}} 
\def \ttxta[#1]{\tilde{\tx}_{{\rm t},#1}} 
\def \ttxda[#1]{\tilde{\tx}_{{\rm d},#1}} 
\def \zta[#1]{z_{{\rm t},#1}}
\def \bzda[#1]{\bz_{{\rm d},#1}}
\def \bzta[#1]{\bz_{{\rm t},#1}}
\def \zda[#1]{z_{{\rm d},#1}}

\def \rxvda[#1]{\rxv_{{\rm d},#1}}

\def \expeq[#1]{\overset{{#1}}{\equiv}}

\def \Enta[#1]{\Omega_{#1}}
\def \vara[#1]{\sigma^2_{#1}}

\def \tauopt{\tau_{{\rm opt}}}
\def \upto {{\nearrow}}
\def \downto {{\searrow}}

\def \EntInnC#1#2#3#4{\EntInn({#2}_{#4}|{#1}_{#3})}

\def \EntInnY#1#2{\EntInn({#1}_{#2})}

\def \AEntInnY#1#2{\bar{\EntInn}({#1}_{#2})}

\def \Pe {P_{\rm e}}
\def \chA {\mathcal{G}}

\def \cIInn#1#2#3{\cI{'}({#1}_{#3};{#2}_{#3})}
\def \cII#1#2#3{\cI{'}({#1};{#2}{#3})}

\def \cIRA#1#2#3{\cI({#1};{#2}{#3})}

\def \chva#1{\chv_{#1}}
\def \chma#1{\chm_{#1}}

\usepackage{amsthm}

\newtheorem{thm}{{Theorem}}
\newtheorem{lem}{Lemma}

\newtheorem{cor}{Corollary}

\newenvironment{thmbis}[1]
  {\renewcommand{\thethm}{\ref{#1}$'$}%
   \addtocounter{thm}{-1}%
   \begin{thm}}
  {\end{thm}}

\begin{document}

\title{Analyzing Training Using Phase Transitions in Entropy---Part I: General Theory}
%\title{On the Computation of Large-Scale Tensor Entropy}
%\author{\IEEEauthorblockN{Kang Gao, Bertrand Hochwald}\\
%\IEEEauthorblockA{Department of Electrical Engineering, %University of Notre Dame, Notre Dame, IN, 46556\\
%Email: \texttt{\{kgao,bhochwald\}@nd.edu}}}
\author{Kang~Gao,~\IEEEmembership{Student Member,~IEEE,} and~Bertrand~M.~Hochwald,~\IEEEmembership{Fellow,~IEEE}%
\thanks{This work was generously supported by NSF Grant \#1731056 and Futurewei Technologies, Inc.}
\thanks{Kang Gao and Bertrand M. Hochwald are with the Department of Electrical Engineering, University of Notre Dame, Notre Dame, IN, 46556 USA (email:kgao@nd.edu; bhochwald@nd.edu).}}
\maketitle
\thispagestyle{plain}
\pagestyle{plain}
\begin{abstract}
We analyze phase transitions in the conditional entropy of a sequence caused by a change in the conditional variables.  Such transitions happen, for example, when training to learn the parameters of a system, since the transition from the training phase to the data phase causes a discontinuous jump in the conditional entropy of the measured system response.  
%We show that the size of the discontinuity is of particular interest in the analysis of the mutual information of systems that use training to learn unknown the parameters.  
For large-scale systems, we present a method of computing a bound on the mutual information obtained with one-shot training, and show that this bound can be calculated using the difference between two derivatives of a conditional entropy.  The system model does not require Gaussianity or linearity in the parameters, and does not require worst-case noise approximations or explicit estimation of any unknown parameters.  The model applies to a broad range of algorithms and methods in communication, signal processing, and machine learning that employ training as part of their operation.
\end{abstract}
\begin{IEEEkeywords}
phase transition, mutual information, training, learning
\end{IEEEkeywords}
\IEEEpeerreviewmaketitle

\section{Introduction}
We begin with a brief synopsis of the results contained herein.  Consider a system model that has input $\tx$ and output $\rx$, where some parameters within the model are unknown.  The system is supplied with known inputs during a ``training phase" to learn the parameters, after which the system is used during its ``data phase".  An analysis of the effects of training is captured by the mutual information between the input and output at the beginning of the data phase, conditioned on the training signals:
\begin{align}
  \lefteqn{\MuI(\tx_{\Tt+1};\rx_{\Tt+1}|\txv_{\Tt},\rxv_{\Tt})} \nonumber\\
 & =\Ent(\rx_{\Tt+1}|\txv_{\Tt},\rxv_{\Tt}) - \Ent(\rx_{\Tt+1}|\txv_{\Tt+1},\rxv_{\Tt}),
  \label{eq:joint_mutual_information}
\end{align}
where $T$ is the number of training symbols, and $\txv_t=[\tx_1,\cdots,\tx_t]^\Tp$ for integer $t$. This quantity measures the amount of information that can be transferred through a single input-output pair in the data phase conditioned on the training.  Generally, this mutual information increases monotonically in $\Tt$ since a greater amount of training allows for better parameter estimates.  We define
\begin{align}
    \cIInn{\txA}{\rxA}{}=\lim_{\Tt\to\infty}\MuI(\tx_{\Tt+1};\rx_{\Tt+1}|\txv_{\Tt},\rxv_{\Tt}),
    \label{eq:MuI_def}
\end{align}
which we assume exists and is finite.

The system model has an input process $\tx_t\in\txA$, and output process $\rx_t\in\rxA$ connected through a joint distribution that has random parameters that are unknown except for what is learned during the training process.  Let
\begin{align*}
&\EntInnC{\txA}{\rxA}{\dx}{\dy}\\
=&\lim_{\Tt\to\infty}\Ent(\rx_{\ceil{(1+\dy)\Tt}+1}|\txv_{\ceil{(1+\dx)\Tt}},\rxv_{\ceil{(1+\dy)\Tt}}),
    \numberthis
    \label{eq:EntInn_cond_def}
\end{align*}
where $\dy\geq -1$ and $\dx\geq -1$, $\ceil{\cdot}$ is the ceiling operation that rounds up to the nearest integer, and we assume that this limit exists.  We also define
\begin{align*}
&\EntInnC{\txA}{\rxA}{\dx^+}{\dy}\\
=&\lim_{\Tt\to\infty}\Ent(\rx_{\ceil{(1+\dy)\Tt}+1}|\txv_{\ceil{(1+\dx)\Tt}+1},\rxv_{\ceil{(1+\dy)\Tt}}),
    \numberthis
    \label{eq:EntInn_cond_def_one_extra}
\end{align*}
where we use $\dx^+$ to denote conditioning on the single extra input $x_{\ceil{(1+\dx)\Tt}+1}$ versus \eqref{eq:EntInn_cond_def}. Note that $\dy=0$ and $\dx=0$ represent the boundary line between the training and data phases.
Then \eqref{eq:joint_mutual_information} and \eqref{eq:MuI_def} yield
\begin{align*}
    \cIInn{\txA}{\rxA}{}=\EntInnC{\txA}{\rxA}{}{} - \EntInnC{\txA}{\rxA}{^+}{},
    \numberthis
    \label{eq:entropy_gap}
\end{align*}
where $\EntInnC{\txA}{\rxA}{}{}$ is defined as $\EntInnC{\txA}{\rxA}{\dx}{\dy}|_{\dy=0,\dx=0}$ and $\EntInnC{\txA}{\rxA}{^+}{}$ is defined as $\EntInnC{\txA}{\rxA}{\dx^+}{\dy}|_{\dy=0,\dx=0}$.

We assume that:
 \begin{align*}
\text{A1:}&\qquad\quad \EntInnC{\txA}{\rxA}{^+}{}  = \lim_{\dy\downto 0} \EntInnC{\txA}{\rxA}{\dy^+}{\dy},
    \numberthis
    \label{eq:training_phase_continue}\\
\text{A2:}&\qquad\quad \EntInnC{\txA}{\rxA}{}{} =\lim_{\dy\downto 0} \EntInnC{\txA}{\rxA}{}{\dy},
    \numberthis
    \label{eq:continue_in_data}
\end{align*}
and then \eqref{eq:entropy_gap} yields
\begin{equation}
    \cIInn{\txA}{\rxA}{} = \lim_{\dy\downto 0} \EntInnC{\txA}{\rxA}{}{\dy} - \lim_{\dy\downto 0} \EntInnC{\txA}{\rxA}{\dy^+}{\dy}.
    \label{eq:Ixy_final}
\end{equation}
We establish (Theorems \ref{thm:derivative_monotonic} \& \ref{thm:integral_thm_monotonic} and Corollary \ref{cor:A3_derivative_relation}) that under certain conditions
\begin{align*}
    \EntInnC{\txA}{\rxA}{\dx}{\dy} =\frac{\partial\EntR(\rxA_\dy|\txA_\dx)}{\partial\dy},
    \numberthis
    \label{eq:derivative_relation1}
\end{align*}
where $\EntR(\rxA_\dy|\txA_\dx)$ is defined as
\begin{equation}
    \EntR(\rxA_\dy|\txA_\dx)=\lim_{\Tt\to\infty}\frac{1}{\Tt}\Ent(\rxv_{\ceil{(1+\dy)\Tt}}|\txv_{\ceil{(1+\dx)\Tt}}),
    \label{eq:EntR_cond_def}
\end{equation}
and we assume that the limit exists and is differentiable with respect to $\dy$.
When $\dx>\dy$, we show that
%If $\tx_t$ is independent of $(\txv_{(1+\dy)\Tt},\rxv_{(1+\dy)\Tt})$, for any $\dx>\dy$, we have
\begin{align*}
    \EntR(\rxA_\dy|\txA_\dx)=\EntR(\rxA_\dy|\txA_\dy),\;\EntInnC{\txA}{\rxA}{\dx}{\dy}=\EntInnC{\txA}{\rxA}{\dy^+}{\dy},
\end{align*}
and \eqref{eq:derivative_relation1} becomes
\begin{align*}
    \EntInnC{\txA}{\rxA}{\dy^+}{\dy} =\frac{\partial\EntR(\rxA_\dy|\txA_\dy)}{\partial\dy}.
    \numberthis
    \label{eq:derivative_relation2}
\end{align*}

We may therefore obtain $\cIInn{\txA}{\rxA}{}$ by computing $\EntInnC{\txA}{\rxA}{}{\dy}$ and $\EntInnC{\txA}{\rxA}{\dy^+}{\dy}$ as derivatives of $\EntR(\rxA_\dy|\txA_\dx)$ (Theorem \ref{thm:computation_of_MuI_from_derivative}).  Finally, in one of our main results, we use $\cIInn{\txA}{\rxA}{}$ to derive a lower bound on the mutual information between the input and output of a system that employs training (Theorem \ref{thm:MI_equal_gap}).  This analysis of training is derived entirely from the derivatives of $\EntR(\rxA_\dy|\txA_\dx)$.

The value of this formulation relies on our ability to find $\EntR(\rxA_\dy|\txA_\dx)$ in a straightforward manner, and we provide guidance on this in Theorem \ref{thm:scale_joint_entropy}.  We make no assumptions of linearity of the system or Gaussianity in any of the processes; nor is it required to form an explicit estimate of the parameters $\chA$ during the training phase. 
Example 5 in Section \ref{thm:computation_of_MuI_from_derivative} applies the theorems to derive the optimal training time in a system with unknown channels. 
The remainder of the paper generalizes the results to high-dimensional systems.  Part II looks at specific applications in communications, signal processing, and machine learning.

\begin{figure}
\includegraphics[width=3.5in]{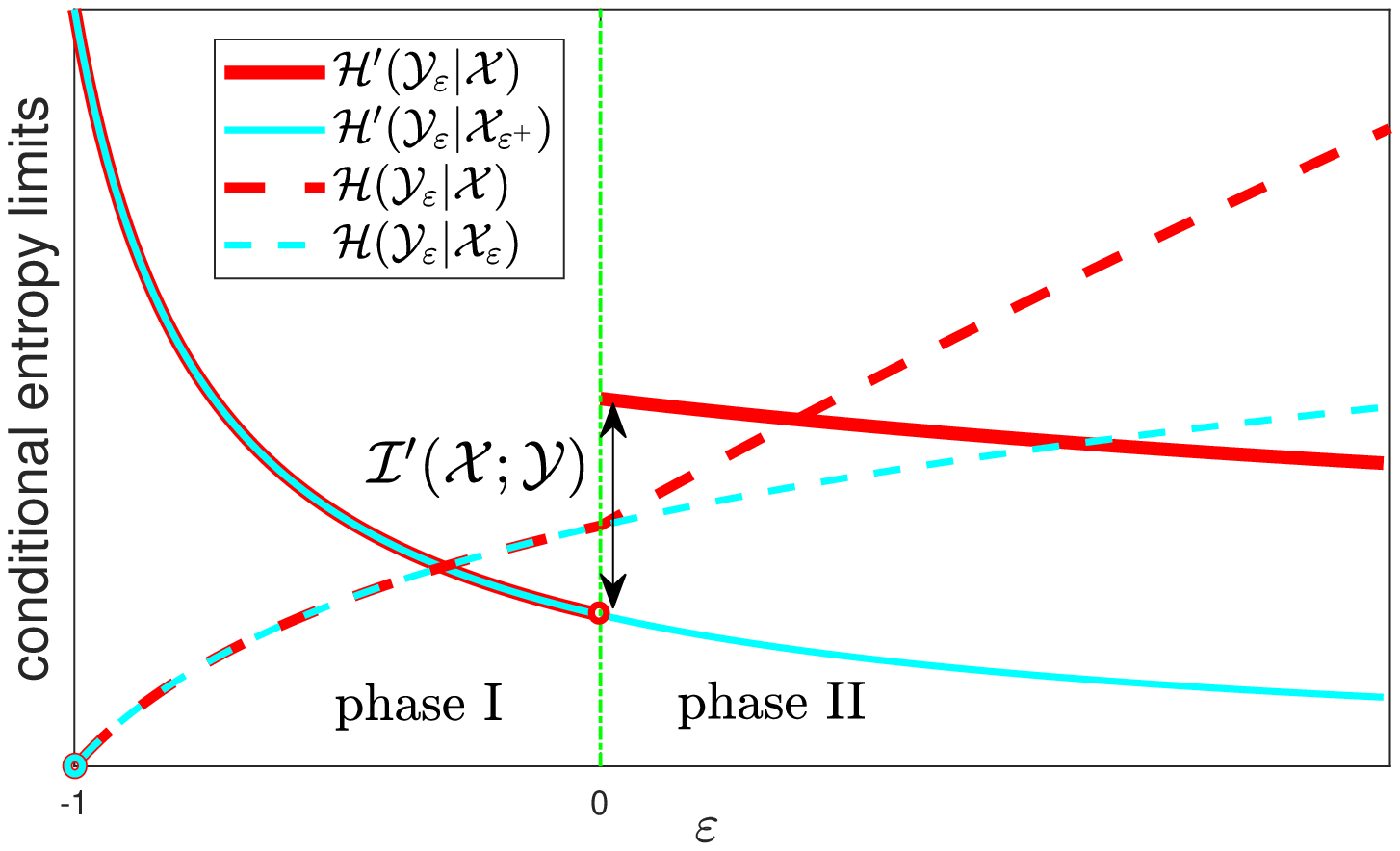}
\centering
    \caption{Qualitative sketch of $\EntInnC{\txA}{\rxA}{}{\dy}$, $\EntInnC{\txA}{\rxA}{\dy^+}{\dy}$, $\EntR(\rxA_\dy|\txA)$ and $\EntR(\rxA_\dy|\txA_\dy)$ as functions of $\dy$ in Phase I $(\dy\leq 0)$ and Phase II $(\dy>0)$ where $\tx_t$ is \iid throughout both phases.  $\EntInnC{\txA}{\rxA}{}{\dy}$ is discontinuous at $\dy=0$ (phase change), while $\EntInnC{\txA}{\rxA}{\dy^+}{\dy}$ is continuous at $0$.  The black arrow shows $\cIInn{\txA}{\rxA}{}$ as the difference between $\EntInnC{\txA}{\rxA}{}{\dy}$ and $\EntInnC{\txA}{\rxA}{\dy^+}{\dy}$.
    Although $\EntR(\rxA_\dy|\txA)$ is not differentiable at $0$, $\EntR(\rxA_\dy|\txA_\dy)$ is.}
    \label{fig:sketch_entropy}
\end{figure}

Qualitative sketches of $\EntInnC{\txA}{\rxA}{}{\dy}$, $\EntInnC{\txA}{\rxA}{\dy^+}{\dy}$, $\EntR(\rxA_\dy|\txA)$ and $\EntR(\rxA_\dy|\txA_\dy)$ are shown in Fig. \ref{fig:sketch_entropy} during phase I (training phase, $\dy\in [-1,0]$) and phase II (data phase, $\dy>0$).  In this example, $\tx_t$ is independent and identically distributed (\iid) throughout both phases; equivalently, the training and data sequences have the same distribution.  The quantity $\EntInnC{\txA}{\rxA}{}{\dy}$ is discontinuous at $\dy=0$ since the input is no longer part of the conditioning for $\dy>0$.  Both  $\EntInnC{\txA}{\rxA}{}{\dy}$ and $\EntInnC{\txA}{\rxA}{\dy^+}{\dy}$ decrease as $\dy$ increases throughout phases I and II. Also shown are $\EntR(\rxA_\dy|\txA)$ and $\EntR(\rxA_\dy|\txA_\dy)$, which are the integrals of $\EntInnC{\txA}{\rxA}{}{\dy}$ and $\EntInnC{\txA}{\rxA}{\dy^+}{\dy}$. Although $\EntR(\rxA_\dy|\txA)$ is not differentiable at $\dy=0$, it is differentiable for  $\dy\neq 0$, while $\EntR(\rxA_\dy|\txA_\dy)$ is differentiable for all $\dy> -1$. 

Phase transitions in entropy have been used in other contexts, including the ``information bottleneck" \cite{wu2020phase}, minimum mean-square error (MMSE) analysis \cite{merhav2010statistical,barbier2018optimal}, and random Boolean networks \cite{lizier2008information}. Here, we focus on the applications to training of phase transitions caused by the change of conditional variables from training signals (with known input) to data signals (with unknown input).
The remainder of the paper is devoted to explanations and justifications of the above assumptions and statements.  We begin by establishing \eqref{eq:derivative_relation1}.

\section{Derivative Relationship between $\EntInn$ and $\EntR$}
\label{sec:monotonic_scalar_process}
Define
\begin{align*}
\EntInn(\rxA_\dy) =\lim_{\Tt\to\infty}\Ent(\rx_{\ceil{(1+\dy)\Tt}+1}|\rxv_{\ceil{(1+\dy)\Tt}}),
    \numberthis
    \label{eq:EntInn_def}
\end{align*}
\begin{equation}
    \EntR(\rxA_\dy)=\lim_{\Tt\to\infty}\frac{1}{\Tt}\Ent(\rxv_{\ceil{(1+\dy)\Tt}}),
    \label{eq:EntR_def}
\end{equation}
which can be considered as  $\EntInnC{\txA}{\rxA}{\dx}{\dy}$ and $\EntR(\rxA_\dy|\txA_{\dx})$ with $\dx=-1$. 
%We omit the conditioning on $\txA$ for the moment.
%\subsection{Derivative relationship between $\EntInn$ and $\EntR$}
We show that, under some conditions, 
 $\EntInn(\rxA_\dy)$ is the derivative of $\EntR(\rxA_\dy)$.
\begin{thm}
\label{thm:derivative_monotonic}
Let $\EntR(\rxA_\dy)$ and its derivative with respect to $\dy$ exist.  Suppose there exists a $\kappa>0$ so that $\Ent(\rx_{t+1}|\rxv_{t})$ is monotonic in $t$ when $t\in[\floor{(1+\dy-\kappa)\Tt},\ceil{(1+\dy+\kappa)\Tt}]$ as $\Tt\to\infty$.  Then
\begin{equation}
    \EntInn(\rxA_\dy) =\frac{\partial\EntR(\rxA_\dy)}{\partial\dy}.
    \numberthis
    \label{eq:derivative_2}
\end{equation}
\end{thm}

\begin{proof}
Without loss of generality, we assume that $\Ent(\rx_{t+1}|\rxv_{t})$ is monotonically decreasing. Using the definition of $\EntR(\rxA_\dy)$ in \eqref{eq:EntR_def}, we have
\begin{align*}
    &\frac{1}{\kappa}(\EntR(\rxA_{\dy+\kappa})-\EntR(\rxA_\dy))\\
    =&\lim_{\Tt\to\infty}\frac{\Ent(\rxv_{\ceil{(1+\dy+\kappa)\Tt}})-\Ent(\rxv_{\ceil{(1+\dy)\Tt}})}{\kappa\Tt}\\
    =&\lim_{\Tt\to\infty}\frac{\sum_{t=\ceil{(1+\dy)\Tt}+1}^{\ceil{(1+\dy+\kappa)\Tt}}\Ent(\rx_{t}|\rxv_{t-1})}{\kappa\Tt}\\
    \leq &\lim_{\Tt\to\infty}\frac{(\kappa\Tt+1)\cdot\Ent(\rx_{\ceil{(1+\dy)\Tt}+1}|\rxv_{\ceil{(1+\dy)\Tt}})}{\kappa\Tt}\\
    =&\lim_{\Tt\to\infty}\Ent(\rx_{\ceil{(1+\dy)\Tt}+1}|\rxv_{\ceil{(1+\dy)\Tt}}).
    \numberthis
    \label{eq:limit_upper_bound}
\end{align*}

Similarly to \eqref{eq:limit_upper_bound}, we also have
\begin{align*}
    &\lim_{\Tt\to\infty}\Ent(\rx_{\ceil{(1+\dy)\Tt}+1}|\rxv_{\ceil{(1+\dy)\Tt}})\\
    \leq&\frac{1}{\kappa}(\EntR(\rxA_\dy)-\EntR(\rxA_{\dy-\kappa})).
    \numberthis
    \label{eq:limit_lower_bound}
\end{align*}
%\begin{align*}
%    &\lim_{\Tt\to\infty}\Ent(\rx_{(1+\dy)\Tt+1}|\rxv_{(1+\dy)\Tt})\\
%    \leq&\frac{1}{\kappa}(\EntR(\rxA_\dy)-\EntR(\rxA_{\dy-\kappa})).
%    \numberthis
%    \label{eq:limit_lower_bound}
%\end{align*}
Let $\kappa\downto 0$ in both \eqref{eq:limit_upper_bound} and \eqref{eq:limit_lower_bound}; because we assume that the derivative of $\EntR(\rxA_\dy)$ exists, these limits both equal this derivative.  Then, the definition of $\EntInn(\rxA_\dy)$ in \eqref{eq:EntInn_def} yields \eqref{eq:derivative_2}.
\end{proof}

An integral equivalent of \eqref{eq:derivative_2} is:
\begin{thm}
\label{thm:integral_thm_monotonic}
For a process $\rxA$ that satisfies:
\begin{enumerate}
    \item there exists a $c>0$ independent of $t$ and $\Tt$ so that $|\Ent(\rx_{t+1}|\rxv_{t})|<c$ for all $t<\Tt$ as $\Tt\to\infty$;
    \item there exists a $\kappa>0$ so that $\Ent(\rx_{t+1}|\rxv_{t})$ is monotonic in $t$ when $t\in[\floor{(1+\dy-\kappa)\Tt},\ceil{(1+\dy+\kappa)\Tt}]$ for all $\dy>-1$, except for a finite number of points as $\Tt\to\infty$;
    \item  $\EntR(\rxA_\dy)$ and $\EntInn(\rxA_\dy)$ exist;
\end{enumerate}
then we have
\begin{equation}
    \EntR(\rxA_\dy) = \int_{-1}^{\dy}\EntInn(\rxA_\dy) d\dy .
    \label{eq:theorem_integral_monotonic_thm}
\end{equation}
\end{thm}
\noindent
The proof is omitted.

Theorems \ref{thm:derivative_monotonic} and \ref{thm:integral_thm_monotonic} are consequences of the entropy chain rule and letting an infinite sum converge to an integral (standard Riemann sum approximation).  Such an analysis has also been used in the context of computing mutual information; for example  \cite{shamai2001impact,guo2005randomly,guo2005mutual,guo2008multiuser,honig2009advances}, where the mutual information between a high-dimensional input vector and a high-dimensional output vector is considered, and the chain rule is applied along the dimension of the input, thus producing a summation of mutual information between a scalar input and the vector output, conditioned on all the previous scalar inputs. In the limit as the dimension goes to infinity, the summation converges to an integral.

Two examples are given.

\noindent
{\em Example 1:} Let $\rxA$ be a stationary process where the joint distribution of any subset of the sequence of random variables is invariant with respect to shifts in the time index \cite{cover2012elements}, and where
\begin{align*}
    \lim_{\Tt\to\infty}\Ent(\rx_{\Tt+1}|\rxv_{\Tt})=\EntR(\rxA)=\lim_{\Tt\to\infty}\frac{1}{\Tt}\Ent(\rxv_T),
\end{align*}
where $\EntR(\rxA)$ is called the ``entropy rate" of $\rxA$. For all $\dy>-1$, we have
\begin{align*}
\EntR(\rxA_\dy)=\lim_{\Tt\to\infty}\frac{1}{\Tt}\Ent(\rxv_{\ceil{(1+\dy)T}})=(1+\dy)\EntR(\rxA),
\end{align*}
and
\begin{equation}
\EntInn(\rxA_\dy)  =  \lim_{\Tt\to\infty}\Ent(\rx_{\ceil{(1+\dy)\Tt}+1}|\rxv_{\ceil{(1+\dy)\Tt}})=\EntR(\rxA).
\end{equation}
  Because of stationarity, $\Ent(\rx_{t+1}|\rxv_t)$ is monotonically decreasing in $t$, and we see that $\EntInn(\rxA_\dy)$ is the derivative of $\EntR(\rxA_\dy)$, as expected.
  
However, we are generally interested in non-stationary processes, and the next example is a simple example containing a ``phase change" at $t=\Tt$.

\noindent
{\em Example 2:} Let
\begin{equation}
    \rx_{t} = \begin{cases} 
      b_t, & t=1,\ldots,\Tt; \\
      b_{(t-1\bmod\Tt)+1}, & t=\Tt+1,\Tt+2\ldots,
      \end{cases}
      \label{eq:repetition_eg}
\end{equation}
where $b_{1},b_{2},\ldots$ are \iid with entropy 1. There are two phases in $\rxA$: the first phase contains \iid elements, while the second phase contains repetitions of the first.  Clearly
\begin{equation}
    \EntInnY{\rxA}{\dy}  = \begin{cases} 
      1, & \dy \in[-1,0); \\
      0, & \dy \geq 0,
      \end{cases}
\end{equation}
and $\Ent(\rx_{t+1}|\rxv_{t})$ is bounded by 1 and is monotonic for all $t$.
Theorem \ref{thm:integral_thm_monotonic} (or inspection) yields
\begin{align*}
    \EntR(\rxA_\dy)= \int_{-1}^{\dy}\EntInnY{\rxA}{u} du=\begin{cases} 
      1+\dy, & \dy<0; \\
      1, & \dy\geq 0.
      \end{cases}
\end{align*}
Note that $\EntR(\rxA_\dy)$ is differentiable everywhere but $\dy=0$.  This point will reappear later.

%Both Theorems \ref{thm:derivative_monotonic} and \ref{thm:integral_thm_monotonic} can be generalized to include conditioning on another process $\txA$. 

%\subsection{Processes where $\EntInnC{\txA}{\rxA}{}{\dy}$ is not the derivative of $\EntR(\rxA_\dy|\txA)$}
\subsection{Processes where $\EntInn(\rxA_\dy)$ is not the derivative of $\EntR(\rxA_\dy)$}
\label{sec:general_scalar_process}
%The previous section establish sufficient conditions where the derivative relationship between $\EntInnY{\rxA}{\gamma}$ and $\EntR(\rxA_\gamma)$ holds, and we focus on triangular processes whose conditional entropy $\Ent(\rx_{t+1}|\rxv_{t})$ is monotonic and bounded; see conditions in Theorems \ref{thm:derivative_monotonic} and \ref{thm:integral_thm_monotonic}. 
We examine, through two examples, what can go wrong when the conditions of Theorems \ref{thm:derivative_monotonic} and \ref{thm:integral_thm_monotonic} are not met.

\subsection*{Example 3: $\Ent(\rx_{t+1}|\rxv_{t})$ oscillates as $t$ increases}
Consider the process
\begin{equation}
    \rxA= (b_1,b_1,b_2,b_2,\ldots,b_{k},b_{k},\ldots),
    \label{eq:oscillation_eg_def}
\end{equation}
where $b_1,b_2,\ldots$ are \iid unit-entropy random variables. Then, for all $t\geq \Tt$,
\begin{equation*}
    \Ent(\rx_{t+1}|\rxv_{t}) = \begin{cases}
      1, \quad  t \quad\text{even} \\
      0, \quad t \quad\text{odd }
   \end{cases}
\end{equation*}
and $\EntInn(\rxA_\dy)$ does not exist for any $\dy$. However, $\EntR(\rxA_\dy)$ exists for all $\dy>-1$ with
\begin{align}
    \EntR(\rxA_\dy|\txA)=\lim_{\Tt\to\infty}\frac{1}{\Tt}\frac{\ceil{(1+\dy)\Tt}}{2}=\frac{1+\dy}{2},
    \label{eq:osc_eg_EntR}
\end{align}
which is differentiable for all $\dy>0$. The conditions for Theorem~\ref{thm:derivative_monotonic} are not met and the derivative relationship \eqref{eq:derivative_2} does not hold.

\subsection*{Example 4: $\Ent(\rx_{t+1}|\rxv_{t})$ is unbounded}
Consider a process $\rxA$ with independent elements whose entropies are
\begin{equation}
    \Ent(\rx_t) = \begin{cases}
      \Tt, & t=\frac{1}{2}\Tt-3\\
      1, & \text{otherwise}
   \end{cases}
   \label{eq:unbounded_eg}
\end{equation}
It is clear that $\Ent(\rx_t|\rxv_{t-1})$ is unbounded at $t=\frac{1}{2}\Tt-3$. Both $\EntInn(\rxA_\dy)$ and $\EntR(\rxA_\dy)$ exist with
\begin{align*}
    \EntInn(\rxA_\dy) = 1,\quad\dy\geq -1,
\end{align*}
\begin{equation}
    \EntR(\rxA_\dy) = \begin{cases}
      1+\dy, & \dy\in(-1,-\frac{1}{2}); \\
      2+\dy, & \dy\geq-\frac{1}{2},
   \end{cases}
\end{equation}
but the conditions for Theorem \ref{thm:integral_thm_monotonic} are not met and the integral relationship \eqref{eq:theorem_integral_monotonic_thm} does not hold everywhere.

Nonetheless, these examples can still be accommodated by expanding the definition of $\EntInn(\rxA_\dy)$.
In Example 3, $\EntInn(\rxA_\dy)$ is not a good representative of $\Ent(\rx_{t+1}|\rxv_{t})$ when $t=\ceil{(1+\dy)\Tt}$ because of its oscillatory behavior. A better representative of $\Ent(\rx_{t+1}|\rxv_{t})$ when $t=\ceil{(1+\dy)\Tt}$ can be found by averaging:
\begin{equation*}
    \AEntInnY{\rxA}{\dy} =\lim_{\kappa\downto 0}\lim_{\Tt\to\infty}\frac{\sum_{t=\ceil{(1+\dy-\frac{\kappa}{2})\Tt}}^{\ceil{(1+\dy+\frac{\kappa}{2})\Tt}-1}\Ent(\rx_{t+1}|\rxv_{t})}{\dy\Tt}.
\end{equation*}
With the new definition, for Example 3, 
\begin{align*}
    \AEntInnY{\rxA}{\dy} =\lim_{\kappa\downto 0}\lim_{\Tt\to\infty}\frac{\kappa\Tt/2}{\kappa\Tt}=\frac{1}{2},
\end{align*}
which is the derivative of $\EntR(\rxA_\dy|\txA)$ shown in \eqref{eq:osc_eg_EntR}. Thus, averaging smooths out the oscillation and expands the class of processes for which Theorem \ref{thm:derivative_monotonic} holds.

In Example 4, $\AEntInnY{\rxA}{\dy}$ is unbounded at $\dy=-\frac{1}{2}$.  By allowing an impulse function in $\AEntInnY{\rxA}{\dy}$ at $\frac{1}{2}$, we may then consider $\EntR(\rxA_\dy)$ as the integral of $\AEntInnY{\rxA}{\dy}$, thereby expanding the class of processes for which Theorem \ref{thm:integral_thm_monotonic} holds.  We do not pursue these issues any further.

\section{Input Process and Mutual Information}
\label{sec:input_process_and_MI}
\subsection{The input process and the parameters}

In the remainder, we consider a triple $(\txA,\rxA,\chA)$, where $\tx_t\in\txA$ and $\rx_t\in\rxA$ are the input and output processes of a system, and $\chA$ is the parameter set.  We assume $t=1,\ldots,\Tb$, where $\Tb$ is the blocklength, defined as the period of time for which the parameters are considered constant. Let $\Tt=\tau\Tb$ be the training time, with $\tau\in(0,1]$, where $\tau$ is the fraction of the total blocklength devoted to learning the unknown parameters through training. The input and output are connected through a conditional distribution parameterized by $\chva{\Tb}\in\chA$, whose value is unknown.  We note that $\chva{\Tb}$ is indexed by $\Tb$, indicating that the parameter set is allowed to grow in size as $\Tb\tendsto\infty$ (and $\Tt=\tau\Tb\tendsto\infty$).  We generally drop the use of $\Tb$, and substitute $\frac{\Tt}{\tau}$ in its place, for a fixed $\tau$. Note that the blocklength $\Tb$ should be an integer, and we may consider $\Tb=\ceil{\frac{\Tt}{\tau}}$ for the fixed $\tau$ as $\Tt$ grows, where the ratio $\frac{\Tt}{\Tb}$ still converges to $\tau$ in the limit. For simplicity in notation, we drop the ceiling notation $\ceil{\cdot}$ and treat $\frac{\Tt}{\tau}$ as an integer.

We make the following assumption:
\begin{align*}
%    \text{A3:}&\quad p(\txv_{(1+\dy)\Tt})=p(\txv_{\Tt})\prod_{t=\Tt+1}^{(1+\dy)\Tt}p(\tx_t),    
    \text{A3:}\;\;& p(\rxv_{\frac{\Tt}{\tau}}|\txv_{\frac{\Tt}{\tau}};\chva{\frac{\Tt}{\tau}}) =\prod_{t=1}^{\frac{\Tt}{\tau}}p(\rx_t|\tx_t;\chva{\frac{\Tt}{\tau}}),
    \numberthis
    \label{eq:fixed_channel_in_data} \\
%   &p(\rxv_{(1+\dy)\Tt}|\txv_{(1+\dy)\Tt};\chva{})=\prod_{t=1}^{(1+\dy)\Tt}p(\rx_t|\tx_t;\chva{}),
   & p(\txv_\frac{\Tt}{\tau})=p(\txv_{\Tt})\prod_{t=\Tt+1}^\frac{\Tt}{\tau}p(\tx_t),
   \numberthis
    \label{eq:iid_data_input_distribution}
\end{align*}
where $p(\rx_t|\tx_t;\chva{\frac{\Tt}{\tau}})$ is a fixed conditional distribution for all $t=1,2,\ldots,\frac{\Tt}{\tau}$ and $p(\tx_t)$ is a fixed distribution for all $t=\Tt+1,\Tt+2,\ldots,\frac{\Tt}{\tau}$. Equation \eqref{eq:fixed_channel_in_data} says that the system is memoryless and time invariant (given the input and parameters) and \eqref{eq:iid_data_input_distribution} says that the input $\tx_t$ is \iid and independent of $\txv_T$ for all $t>\Tt$.  The distributions of $\tx_t$ during training and afterward can therefore differ.  We use the common convention of writing $p(\txv_{\Tt})$ and $p(\tx_t)$ when we mean $p_{\txv_{\Tt}}(\cdot)$ and $p_{\tx_t}(\cdot)$, even though these functions can differ. 
Under A3, the distributions of $(\txA,\rxA,\chA)$ are described by the set of known distributions
\begin{equation}
\cP(\Tt,\tau)=\{p(\rx|\tx;\chva{\frac{\Tt}{\tau}}),p(\chva{\frac{\Tt}{\tau}}),p(\txv_\Tt),p(\tx_{\Tt+1})\},
\label{eq:P_set}
\end{equation}
which depends on $\tau$.  These distributions are used to calculate all of the entropies and mutual informations throughout, and hence, these quantities may depend on $\tau$ and can be thought of as ``ergodic" in the sense that they average over realizations of $\chv_{\frac{\Tt}{\tau}}$.

Both Theorems \ref{thm:derivative_monotonic} and \ref{thm:integral_thm_monotonic} can be generalized to include conditioning on $\txA$, thus leading to the following corollary, provided that $\EntR(\rxA_\dy|\txA_\dx)$ and its derivative with respect to $\dy$ exist.
\begin{cor}
\label{cor:A3_derivative_relation}
Under Assumption A3, for $\dy>0$,
\begin{align*}
    \EntInnC{\txA}{\rxA}{}{\dy} =\frac{\partial\EntR(\rxA_\dy|\txA)}{\partial\dy},
    \numberthis
    \label{eq:derivative_data}
\end{align*}
\begin{align*}
    \EntInnC{\txA}{\rxA}{\dy^+}{\dy} =\frac{\partial\EntR(\rxA_\dy|\txA_\dy)}{\partial\dy}.
    \numberthis
    \label{eq:derivative_training}
\end{align*}
If $\tx_t$ are \iid for all $t$, then we have
\begin{align*}
    \EntInnC{\txA}{\rxA}{\dx}{\dy} =\frac{\partial\EntR(\rxA_\dy|\txA_\dx)}{\partial\dy}
    \numberthis
    \label{eq:derivative_more_iid_training}
\end{align*}
for all $\dy,\dx>-1$ and $\dy\neq\dx$. Also,
\begin{align*}
    \EntInnC{\txA}{\rxA}{^+}{} =\left.\frac{\partial\EntR(\rxA_\dy|\txA_\dy)}{\partial\dy}\right|_{\dy=0}.
    \numberthis
    \label{eq:derivative_training_0}
\end{align*}
\end{cor}

\begin{proof}
Under Assumption A3, for all $\dx\geq 0$, we have
\begin{align*}
    \Ent(\rx_{t+1}|\txv_{\ceil{(1+\dx)\Tt}},&\rxv_{t})\leq\Ent(\rx_{t+1}|\txv_{\ceil{(1+\dx)\Tt}},\rxv_{t-1})\\
    =&\;\Ent(\rx_{t}|\txv_{\ceil{(1+\dx)\Tt}},\rxv_{t-1}),
    \numberthis
    \label{eq:monotonic_A3}
\end{align*}
when $\Tt+1\leq t\leq \ceil{(1+\dx)\Tt}-1$ or $t\geq \ceil{(1+\dx)\Tt}+1$. Here, we use that the input is \iid and the system is memoryless and time invariant; the inequality follows from the fact that conditioning reduces entropy.  

Therefore, $\forall \kappa\in(0,\dy)$, $\Ent(\rx_{t+1}|\txv_{\Tt},\rxv_{t})$ is monotonic decreasing in $t$ for $t\in[\floor{(1+\dy-\kappa)\Tt},\ceil{(1+\dy+\kappa)\Tt}]$ when $\Tt>\frac{1}{\dy-\kappa}$. Then, Theorem \ref{thm:derivative_monotonic} yields \eqref{eq:derivative_data}.

Also, $\forall \kappa\in(0,\dy),\dx>2\dy$, $\Ent(\rx_{t+1}|\txv_{\ceil{(1+\dx)\Tt}},\rxv_{t})$ is monotonic decreasing in $t$ for $t\in[\floor{(1+\dy-\kappa)\Tt},\ceil{(1+\dy+\kappa)\Tt}]$ when $\Tt>\max(\frac{1}{\dy-\kappa},\frac{2}{\dy-\dx-\kappa})$. Then, Theorem \ref{thm:derivative_monotonic} yields
\begin{equation}
    \EntInnC{\txA}{\rxA}{\dx}{\dy} =\frac{\partial\EntR(\rxA_\dy|\txA_{\dx})}{\partial\dy}.
    \label{eq:derivative_1}
\end{equation}
Assumption A3 yields
\begin{align}
    \EntInnC{\txA}{\rxA}{\dx}{\dy}=\EntInnC{\txA}{\rxA}{\dy^+}{\dy},
    \label{eq:H_prime_equ}
\end{align}
where $\EntInnC{\txA}{\rxA}{\dy^+}{\dy}$ is defined in \eqref{eq:EntInn_cond_def_one_extra}, and 
\begin{align}
    \EntR(\rxA_\dy|\txA_{\dx})=\EntR(\rxA_\dy|\txA_{\dy}).
    \label{eq:H_equ}
\end{align}
Therefore, \eqref{eq:derivative_1} becomes \eqref{eq:derivative_training}.

If $\tx_t$ are \iid for all $t$, then \eqref{eq:monotonic_A3} is valid for all $t\leq \ceil{(1+\dx)\Tt}-1$ or $t\geq \ceil{(1+\dx)\Tt}+1$. Therefore, Theorem \ref{thm:derivative_monotonic} yields \eqref{eq:derivative_more_iid_training}. By taking $\dy=0$ and $\dx>0$, \eqref{eq:derivative_more_iid_training}, \eqref{eq:H_prime_equ}, and \eqref{eq:H_equ} then yield \eqref{eq:derivative_training_0}.
\end{proof}

\subsection{Computation of $\cIInn{\txA}{\rxA}{}$ and significance of assumptions}
We defer our applications of $\cIInn{\txA}{\rxA}{}$ until Section \ref{sec:MuI_analysis}, but now show it may readily be computed with the help of Corollary \ref{cor:A3_derivative_relation}.  It is shown in \eqref{eq:entropy_gap} that $\cIInn{\txA}{\rxA}{}$ can be computed as the difference between $\EntInnC{\txA}{\rxA}{}{}$ and $\EntInnC{\txA}{\rxA}{^+}{}$. These quantities can be computed as derivatives of $\EntR(\rxA_\dy|\txA)$ and $\EntR(\rxA_\dy|\txA_\dy)$, provided that these limits exist and are differentiable.

\begin{thm}
\label{thm:computation_of_MuI_from_derivative}
 Under Assumptions A1--A3, we have
\begin{equation}
    \cIInn{\txA}{\rxA}{} = \lim_{\dy\downto 0}\frac{\partial\EntR(\rxA_\dy|\txA)}{\partial\dy} - \lim_{\dy\downto 0} \frac{\partial\EntR(\rxA_\dy|\txA_\dy)}{\partial\dy}.
    \label{eq:MuI_from_derivative}
\end{equation}
\end{thm}
\begin{proof}
Corollary \ref{cor:A3_derivative_relation} and \eqref{eq:Ixy_final} yield \eqref{eq:MuI_from_derivative}.
\end{proof}

Assumptions A1--A3 in \eqref{eq:training_phase_continue}, \eqref{eq:continue_in_data}, \eqref{eq:fixed_channel_in_data}, and \eqref{eq:iid_data_input_distribution}, are important for Theorem \ref{thm:computation_of_MuI_from_derivative}.
A3 is often met in practice for a memoryless and time-invariant system with \iid input in the data phase, independent of the input and output during training. However, we do not have a complete characterization of the processes $\txA$ and $\rxA$ that meet Assumptions A1 and A2.
Even though we cannot characterize these processes, A1 and A2 may be verified on a case-by-case basis by examining expressions of  $\EntInnC{\txA}{\rxA}{\dy^+}{\dy}$ and $\EntInnC{\txA}{\rxA}{}{\dy}$ with $\dy\geq 0$ using Corollary \ref{cor:A3_derivative_relation}; see Example 5. 
%Assumption A2 requires knowledge of $\EntInnC{\txA}{\rxA}{}{}$ that we cannot be obtained from Corollary \ref{cor:A3_derivative_relation}. 
The following lemma then helps to verify A2 using A3.

\begin{lem}
\label{lem:bound_for_A2}
If Assumption A3 is met and
\begin{align}
    \lim_{\dy\downto 0}\EntInnC{\txA}{\rxA}{}{\dy}=\lim_{\dx\upto 0}\EntInnC{\txA}{\rxA}{\dx}{},
    \label{eq:two_bound_met}
\end{align}
then Assumption A2 is met.
\end{lem}
\begin{proof}
Under Assumption A3, for $t\geq\Tt+1$, we have
\begin{align*}
  \Ent(\rx_{t+1}|\txv_{\Tt},\rxv_{t}) \leq\Ent(\rx_t|\txv_{\Tt},\rxv_{t-1})
    \numberthis
    \label{eq:monotonic_condition_2}
\end{align*}
which is also shown in \eqref{eq:monotonic_A3}.
Therefore,
\begin{equation*}
   \lim_{\dy\downto 0}\EntInnC{\txA}{\rxA}{}{\dy}\leq \EntInnC{\txA}{\rxA}{}{}.
\end{equation*}
Conditioning to reduce entropy again yields
\begin{align*}
  \Ent(\rx_{\Tt+1}|\txv_{\Tt},\rxv_{\Tt}) \leq&\; \Ent(\rx_{\Tt+1}|\txv_{\ceil{(1-\dx)\Tt}},\rxv_{\Tt}),
\end{align*}
for any $\dx>0$ and therefore,
\begin{equation*}
    \EntInnC{\txA}{\rxA}{}{}\leq \lim_{\dx\upto 0}\EntInnC{\txA}{\rxA}{\dx}{}.
\end{equation*}
Equation \eqref{eq:two_bound_met} then implies A2.
\end{proof}
This lemma is helpful because it replaces computing $\EntInnC{\txA}{\rxA}{}{}$ with computing
$\EntInnC{\txA}{\rxA}{}{\dy}$ and $\EntInnC{\txA}{\rxA}{\dx}{}$, which can be done using Corollary \ref{cor:A3_derivative_relation}. 

\begin{figure} 
\includegraphics[width=3.5in]{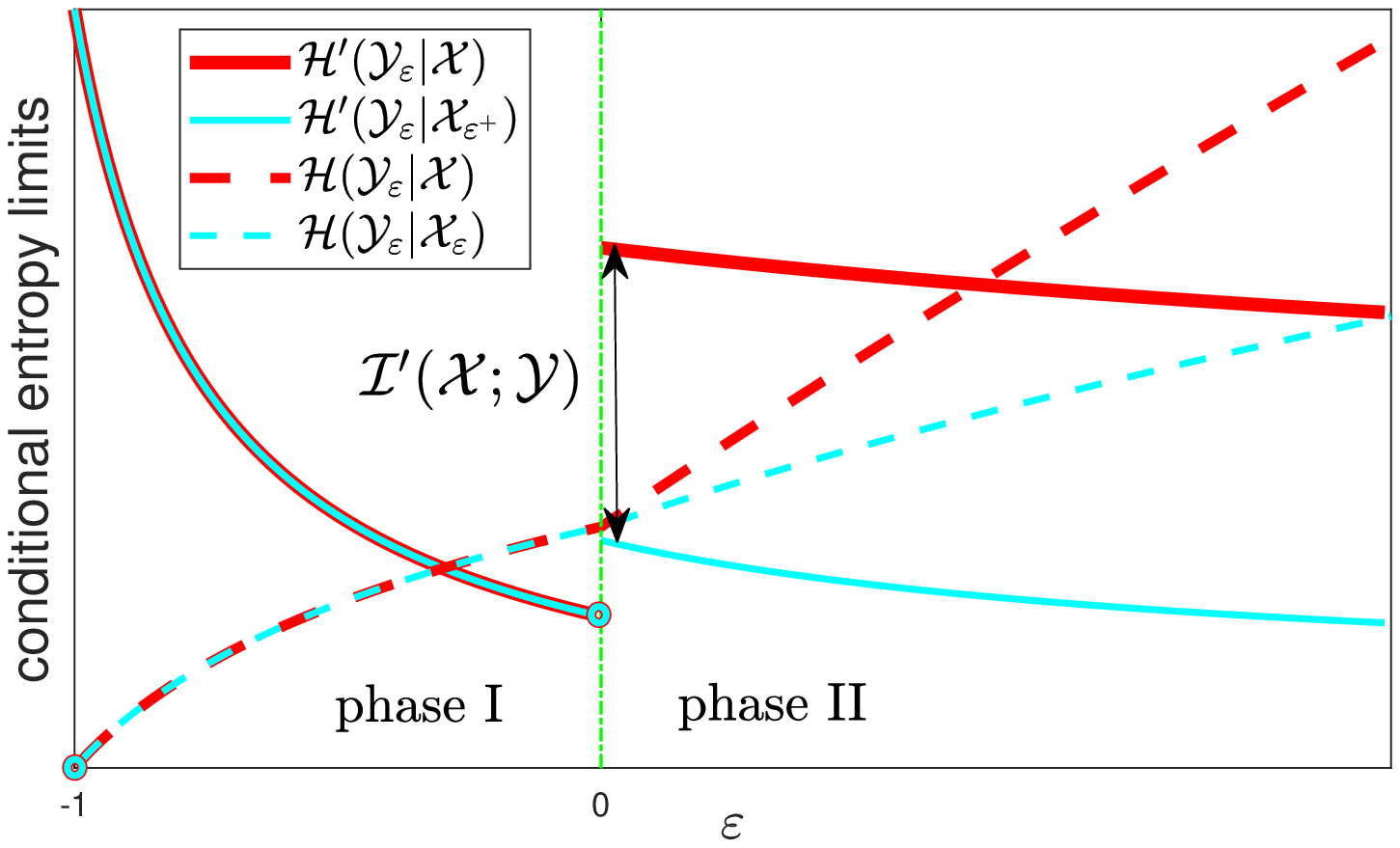}
\centering
    \caption{Similar to Fig.\ \ref{fig:sketch_entropy}
    except the input distributions in Phases I and II differ. Both $\EntInnC{\txA}{\rxA}{}{\dy}$ and $\EntInnC{\txA}{\rxA}{\dy^+}{\dy}$ are discontinuous at $0$.  Their difference (black arrow) is $\cIInn{\txA}{\rxA}{}$. Both  $\EntR(\rxA_\dy|\txA)$ and $\EntR(\rxA_\dy|\txA_\dy)$ are not differentiable at 0.}
    \label{fig:sketch_entropy_non_iid}
\end{figure}

A qualitative sketch that illustrates Corollary \ref{cor:A3_derivative_relation} and Theorem \ref{thm:computation_of_MuI_from_derivative} is shown in Fig. \ref{fig:sketch_entropy}, where $\tx_t$ is \iid throughout both phases, and $\EntInnC{\txA}{\rxA}{\dy^+}{\dy}$ is continuous at $\dy=0$ (phase change). The entropies $\EntInnC{\txA}{\rxA}{}{\dy}$ and $\EntInnC{\txA}{\rxA}{\dy^+}{\dy}$ are the derivatives of $\EntR(\rxA_\dy|\txA)$ and $\EntR(\rxA_\dy|\txA_\dy)$ (Corollary \ref{cor:A3_derivative_relation}) and the mutual information $\cIInn{\txA}{\rxA}{}$ can be computed as the difference between these derivatives (Theorem \ref{thm:computation_of_MuI_from_derivative}). When the input distribution in the training phase (phase I) differs from that in the data phase (phase II), the corresponding sketch is shown in Fig. \ref{fig:sketch_entropy_non_iid}, which is similar to Fig.\ \ref{fig:sketch_entropy}, except that $\EntInnC{\txA}{\rxA}{\dy^+}{\dy}$ is discontinuous at $\dy=0$ because of the changing input distribution. 

\subsection{Mutual information bounds}
\label{sec:MuI_analysis}
%Because of A3, $\cII{\txA}{\rxA}{}$ defined in \eqref{eq:MuI_def} is a function of $\tau$.  
We generalize the definition of $\cII{\txA}{\rxA}{}$ to $\cII{\txA_{\dy}}{\rxA_{\dy}}{}$ with $\dy\in[-1,\frac{1}{\tau}-1)$ as
\begin{align*}
  \cII{\txA_\dy}{\rxA_\dy}{}=  &\lim_{\Tt\to\infty}\MuI(\tx_{\ceil{(1+\dy)\Tt}+1};\\
 &\rx_{\ceil{(1+\dy)\Tt}+1}|\txv_{\ceil{(1+\dy)\Tt}},\rxv_{\ceil{(1+\dy)\Tt}}),
  \numberthis
  \label{eq:cI_def_with_P_tau}
\end{align*}
and we define
\begin{align*}
    \cIRA{\txA_\dy}{\rxA_\dy}{}=&\lim_{\Tt\to\infty}\frac{\tau}{\Tt}\MuI(\txv^{\ceil{(1+\dy)\Tt}+1};\\
    &\rxv^{\ceil{(1+\dy)\Tt}+1}|\txv_{\ceil{(1+\dy)\Tt}},\rxv_{\ceil{(1+\dy)\Tt}}),
    \numberthis
    \label{eq:joint_MuI_thm}
\end{align*}
where $\txv^t=[\tx_t,\tx_{t+1},\cdots,\tx_{\frac{\Tt}{\tau}}]$. Both $\cII{\txA_\dy}{\rxA_\dy}{}$ and $\cIRA{\txA_\dy}{\rxA_\dy}{}$ are computed using $\cP(\Tt,\tau)$ defined in \eqref{eq:P_set}, and are therefore functions of $\tau$.
When $\dy=-1$, $\cIRA{\txA_{-1}}{\rxA_{-1}}{}$ is the mutual information without any training. We show in the following theorem that
both $\cII{\txA_\dy}{\rxA_\dy}{}$ and $\cIRA{\txA_\dy}{\rxA_\dy}{}$ have close relationship with $\cII{\txA}{\rxA}{}$ provided that these limits exist:

%we provide a computable lower bound for the mutual information between the output and input in the data phase conditioned on the training signals. 
%\subsection{Mutual information and inequalities}

\begin{thm}
\label{thm:MI_equal_gap}
Under A3, for all $\dy_1,\dy_2\geq 0$, $\dy_1>\dy_2$, $\tau\in(0,1)$, 
\begin{align*}
\cII{\txA_{\dy_1}}{\rxA_{\dy_1}}{}\geq \cII{\txA_{\dy_2}}{\rxA_{\dy_2}}{},
    \numberthis
    \label{eq:monotonic_MuI}
\end{align*}
and,
\begin{align*}
    &\cIRA{\txA_{-1}}{\rxA_{-1}}{}\geq\;\cIRA{\txA}{\rxA}{}
    \numberthis
    \label{eq:joint_MuI_expression}\\
   & \qquad\geq\; \int_{0}^{\frac{1}{\tau}-1}\tau \cII{\txA_{\dy}}{\rxA_{\dy}}{}d\dy 
    \numberthis
    \label{eq:MuI_int_expression}\\
    &\qquad\geq\;(1-\tau)\cII{\txA}{\rxA}{}
    \numberthis
    \label{eq:MuI_begin_lower_bound}\\
    &\qquad\geq\; (1-\tau)\lim_{T\to\infty}\MuI(\tx_{T+1};\rx_{ T+1}|\hat{\chv}_{\frac{\Tt}{\tau}}),
    \numberthis
    \label{eq:MuI_estimate_lower_bound}
\end{align*}
where $\hat{\chv}_{\frac{\Tt}{\tau}}$ is any estimate of $\chva{\frac{\Tt}{\tau}}$ that is a function of $(\txv_{ T},\rxv_{ T})$.
%and $\cIInn{\txA}{\rxA}{\gamma=1}$ is defined as
%\begin{align*}
%    \cIInn{\txA}{\rxA}{}=\lim_{\gamma\upto 1}\cIInn{\txA}{\rxA}{\gamma}.
%\end{align*}
\end{thm}

\begin{proof}
Assumption A3 implies that $\tx_t$ is independent of $(\tx_k,\rx_k)$ when $k\neq t$ and $t\geq \Tt+1$. Therefore, for all $t\geq \Tt+1$,
\begin{align*}
    &\MuI(\tx_{t+1};\rx_{t+1}|\txv_{t},\rxv_{t}) - \MuI(\tx_{t};\rx_{t}|\txv_{t-1},\rxv_{t-1})\\
    \stackrel{\rm{( a)}}{=}&(\Ent(\tx_{t+1}) - \Ent(\tx_{t+1}|\txv_t,\rxv_{t+1})) \\
    &\qquad - (\Ent(\tx_{t}) - \Ent(\tx_{t}|\txv_{t-1},\rxv_{t}))\\
    =&\Ent(\tx_{t}|\txv_{t-1},\rxv_{t}) - \Ent(\tx_{t+1}|\txv_t,\rxv_{t+1})\\
    \stackrel{\rm{( b)}}{=}&\Ent(\tx_{t+1}|\txv_{t-1},\rxv_{t-1},\rx_{t+1}) - \Ent(\tx_{t+1}|\txv_t,\rxv_{t+1})\\
    \stackrel{\rm{( c)}}{\geq}& 0.
    \numberthis
    \label{eq:monotonic_MuI_T}
\end{align*}
Here, $^{(a)}$ uses the independence between $\tx_t$ and $(\txv_{t-1},\rxv_{t-1})$, $^{(b)}$ uses Assumption A3, $^{(c)}$ uses conditioning to reduce entropy. Thus, $\MuI(\tx_t;\rx_t|\txv_{t-1},\rxv_{t-1})$ is monotonically increasing with $t$ for all $t>\Tt$. In the limit when $\Tt\to\infty$, \eqref{eq:monotonic_MuI_T} yields \eqref{eq:monotonic_MuI}.  A sufficient condition for achieving equality in $^{(c)}$ is when $\chva{\frac{\Tt}{\tau}}$ can be estimated perfectly from $(\txv_{\Tt},\rxv_{\Tt})$, and both entropies on the right hand side of $^{(b)}$ equal $\Ent(\tx_{t+1}|\chva{\frac{\Tt}{\tau}},\rx_{t+1})$. 
%All the mutual informations and entropies are computed using the distribution set $\cP(\Tt,\tau)$, which is not expressed explicitly for simplicity in notation.

Moreover,
\begin{align*}
    &\MuI(\txv^{\Tt+1};\rxv_{\frac{\Tt}{\tau}}|\txv_\Tt)\\
    =&\MuI(\txv^{\Tt+1};\rxv_{\Tt}|\txv_\Tt)+\MuI(\txv^{\Tt+1};\rxv^{\Tt+1}|\txv_\Tt,\rxv_{\Tt})\\
    =&\MuI(\txv^{\Tt+1};\rxv^{\Tt+1}|\txv_\Tt,\rxv_{\Tt}),
\end{align*}
where the first equality uses the chain rule and the second uses that $\txv^{\Tt+1}$ is independent of $(\txv_\Tt,\rxv_\Tt)$. Therefore,
\begin{align*}
    &\MuI(\txv^{\Tt+1};\rxv^{\Tt+1}|\txv_\Tt,\rxv_{\Tt}) = \Ent(\rxv_{\frac{\Tt}{\tau}}|\txv_{\Tt})-\Ent(\rxv_{\frac{\Tt}{\tau}}|\txv_{\frac{\Tt}{\tau}})\\
    &\leq\Ent(\rxv_{\frac{\Tt}{\tau}})-\Ent(\rxv_{\frac{\Tt}{\tau}}|\txv_{\frac{\Tt}{\tau}})=\MuI(\txv_{\frac{\Tt}{\tau}};\rxv_{\frac{\Tt}{\tau}}),
\end{align*}
which, with \eqref{eq:joint_MuI_thm}, yields \eqref{eq:joint_MuI_expression}.

To prove \eqref{eq:MuI_int_expression}, we first show following inequality:
\begin{align*}
    &\MuI(\txv^{\Tt+1};\rxv^{\Tt+1}|\txv_{\Tt},\rxv_{\Tt})\\
    \stackrel{}{=}&\;\Ent(\txv^{\Tt+1}|\txv_{\Tt},\rxv_{\Tt}) - \Ent(\txv^{\Tt+1}|\txv_{\Tt},\rxv_{\frac{\Tt}{\tau}})\\
    \stackrel{\rm{( a)}}{=}&\sum_{t={\Tt+1}}^{\frac{\Tt}{\tau}}(\Ent(\tx_t|\txv_{t-1},\rxv_{\Tt}) - \Ent(\tx_t|\txv_{t-1},\rxv_{\frac{\Tt}{\tau}}))\\
    \stackrel{\rm{( b)}}{\geq}&\sum_{t={\Tt+1}}^{\frac{\Tt}{\tau}}(\Ent(\tx_t|\txv_{t-1},\rxv_{t-1}) - \Ent(\tx_t|\txv_{t-1},\rxv_{t}))\\
    =& \sum_{t=\Tt+1}^{\frac{\Tt}{\tau}}\MuI(\tx_t;\rx_t|\txv_{t-1},\rxv_{t-1}).
    \numberthis
    \label{eq:chain_rule_MuI_bound}
\end{align*}
Here, $^{(a)}$ uses the chain rule, $^{(b)}$ uses conditioning to reduce entropy. Equality in $^{(b)}$ can be achieved when $\chva{\frac{\Tt}{\tau}}$ is estimated perfectly from $(\txv_{\Tt},\rxv_{\Tt})$.  Then, by normalizing \eqref{eq:chain_rule_MuI_bound} by $\frac{\tau}{\Tt}$ and letting $\Tt\to\infty$, we see that the sum converges to the integral, which proves \eqref{eq:MuI_int_expression}. Then, \eqref{eq:monotonic_MuI} yields \eqref{eq:MuI_begin_lower_bound}. 
 
 Also, 
 \begin{align*}
     &\MuI(\tx_{\Tt+1};\rx_{\Tt+1}|\txv_{\Tt},\rxv_{\Tt})\\
     =&\Ent(\tx_{\Tt+1})-\Ent(\tx_{\Tt+1}|\txv_{\Tt},\rxv_{\Tt},\rx_{\Tt+1})\\
     \stackrel{\rm{( a)}}{=}&\Ent(\tx_{\Tt+1})-\Ent(\tx_{\Tt+1}|\txv_{\Tt},\rxv_{\Tt},\hat{\chv}_{\frac{\Tt}{\tau}},\rx_{\Tt+1})\\
     \stackrel{\rm{( b)}}{\geq}&\Ent(\tx_{\Tt+1})-\Ent(\tx_{\Tt+1}|\hat{\chv}_{\frac{\Tt}{\tau}},\rx_{\Tt+1})\\
     =&\MuI(\tx_{\Tt+1};\rx_{\Tt+1}|\hat{\chv}_{\frac{\Tt}{\tau}}),
 \end{align*}
 where $^{(a)}$ uses that $\hat{\chv}_{\frac{\Tt}{\tau}}$ is a function of $(\txv_{\Tt},\rxv_{\Tt})$, and $^{(b)}$ uses conditioning to reduce entropy. By taking the limit $\Tt\to \infty$, we have \eqref{eq:MuI_estimate_lower_bound}.
\end{proof}

%In this theorem, $\cIRA{\txA_{-1}}{\rxA_{-1}}{}$ is the mutual information between the input and output without training; $\cIRA{\txA}{\rxA}{}$ is the mutual information between the input and output in the entire data phase only; $\cII{\txA_{\dy}}{\rxA_{\dy}}{}$ is the mutual information between a single input and output immediately after $(1+\dy)\Tt$ input and output; $\MuI(\tx_{\Tt+1};\rx_{\Tt+1}|\hat{\chv}_{\frac{\Tt}{\tau}})$ is the mutual information between a single input and output with an explicit estimate of ${\chv}_{\frac{\Tt}{\tau}}$.

\subsection{Discussion of lower bounds}

The quantities in \eqref{eq:joint_MuI_expression}--\eqref{eq:MuI_estimate_lower_bound} have a long history of being studied in various contexts. For example, $\cIRA{\txA_{-1}}{\rxA_{-1}}{}$ describes the mutual information between the entire input and output block. This quantity is studied in \cite{marzetta1999capacity,hochwald2000unitary,zheng2002communication} in the context of a linear system with additive Gaussian noise and vector inputs and outputs; analytical expressions are obtained in special cases when the blocklength is larger than the dimension of the input, or the noise power is small. 

The right-hand side of \eqref{eq:joint_MuI_expression} $\cIRA{\txA}{\rxA}{}$ describes the mutual information between the input and output in the data phase conditioned on $\Tt$ training symbols, and the right-hand side of \eqref{eq:MuI_int_expression} integrates the mutual information between each input and output pair in the data phase where the training and data output together are used to refine the estimate of $\chva{\frac{\Tt}{\tau}}$. In \cite{takeuchi2012large, takeuchi2013achievable}, linear systems with additive Gaussian noise are considered, starting with $\cIRA{\txA}{\rxA}{}$, which is then lower bounded by an integral similar to the right-hand side of \eqref{eq:MuI_int_expression}.  This integral is further lower-bounded by obtaining a linear minimum mean-squared error (LMMSE) estimate of $\chva{\frac{\Tt}{\tau}}$. This lower bound is maximized when $\tau=0$ since any amount of training at the expense of data is harmful to the total throughput for the block; the unknown parameters can be jointly estimated with the data without explicit training signals.  In general, when the input is \iid during the whole block, 
\eqref{eq:joint_MuI_expression} and \eqref{eq:MuI_int_expression} are maximized when $\tau=0$.

In contrast, the right-hand sides of \eqref{eq:MuI_begin_lower_bound} and \eqref{eq:MuI_estimate_lower_bound} assume $\chva{\frac{\Tt}{\tau}}$ is learned only through the training phase (and not through the data phase)---so called ``one-shot" learning, which is our primary interest. In particular, \eqref{eq:MuI_begin_lower_bound} does not form an explicit estimate of $\chva{\frac{\Tt}{\tau}}$, and is an upper bound on \eqref{eq:MuI_estimate_lower_bound}, which forms an explicit estimate derived only from the training. Special cases of one-shot learning with explicit estimates of system parameters are analyzed in \cite{hassibi2003much,li2016much,li2017channel}.  A linear system with additive Gaussian noise is analyzed in \cite{hassibi2003much}, where the system parameters are estimated in the training phase, and the worst-case noise analysis then produces a lower bound. The worst-case noise analysis developed in \cite{hassibi2003much} treats the estimation error as additive Gaussian noise.  In \cite{li2016much,li2017channel}, systems with additive Gaussian noise and one-bit quantization at the output are considered, where the Bussgang decomposition is used to reformulate the nonlinear quantizer as a linear function with additional noise, followed by a worst-case noise analysis. Training times that maximize these lower bounds are generally nonzero since only the training phase is used to learn the unknown parameters, and further refinement of these parameters during the data phase is not performed.

Theorem \ref{thm:MI_equal_gap} does not require linearity in the system parameters or Gaussian additive noise, and therefore has wide potential applicability to analyzing systems that use training.  The optimum amount of training is found from \eqref{eq:MuI_begin_lower_bound}:
\begin{align}
    \tauopt=\argmax_{\tau}(1-\tau)\cII{\txA}{\rxA}{}
    \label{eq:opt_training_tau}
\end{align}
and the corresponding rate per receive symbol is
\begin{equation}
    \Ropt = (1-\tauopt)\cII{\txA}{\rxA}{}\bigr|_{\tau=\tauopt}.
    \label{eq:Ropt}
\end{equation}
As shown in \eqref{eq:MuI_estimate_lower_bound}, this is an upper bound on the rate achievable by any method that estimates the unknown parameters from the training data.  According to Theorem \ref{thm:computation_of_MuI_from_derivative}, derivatives of $\EntR(\rxA_\dy|\txA_\dx)$, which can be obtained from the single entropy function $\EntR(\rxA_\dy|\txA)$, are needed to compute this quantity.  This is the subject of the next section.

\section{Computation of $\EntR(\rxA_\dy|\txA_\dx)$ and an example}

%\subsection{Availability of $\EntR(\rxA_\dy|\txA)$}
%\label{subsec:compute_EntR}
%Theorem \ref{thm:computation_of_MuI_from_derivative} shows that derivatives of $\EntR(\rxA_\dy|\txA_\dx)$ can be used to compute mutual information in the large $\Tt$ limit.  
Closed-form expressions for $\EntR(\rxA_\dy|\txA_\dx)$ may be derived from $\EntR(\rxA_\dy|\txA)$, when this is available.  For example, in some cases $\EntR(\rxA_\dy|\txA)$ can be obtained through methods employed in statistical mechanics by treating the conditional entropy as free energy in a large-scale system.  Free energy is a fundamental quantity \cite{castellani2005spin,mezard2009information} that has been analyzed through the powerful ``replica method", and this, in turn, has been applied to entropy calculations in 
%compressed sensing \cite{sakata2013sample,schulke2016phase,kabashima2016phase},  
machine learning \cite{engel2001statistical,opper1996statistical,shinzato2008learning,ha1993generalization} and wireless communications \cite{wen2015performance,wen2015joint,wen2016bayes}, in both linear and nonlinear systems.  
%These references generally consider vector inputs and outputs, and our results generalize to these cases as shown in Section \ref{sec:high_dimen_generalize}.

The entropy $\EntR(\rxA|\txA)$ (equivalent to $\dy=0$) is considered in \cite{engel2001statistical,opper1996statistical,ha1993generalization,shinzato2008learning}, where the input is multiplied by an unknown vector as an inner product and the passes through a nonlinearity to generate a scalar output. In \cite{engel2001statistical,opper1996statistical,ha1993generalization}, the input are \iid, while orthogonal inputs are considered in \cite{shinzato2008learning}. 
%In \cite{engel2001statistical,opper1996statistical,shinzato2008learning}, the output is the sign of the input, while a sigmoid nonlinearity with a continuous output is considered in \cite{ha1993generalization}. 
The entropy $\EntR(\rxA_{\dy}|\txA)$ for MIMO systems is considered in \cite{wen2015performance,wen2015joint,wen2016bayes}, where the inputs are \iid in the training phase and are \iid in the data phase, but the distributions in the two phases can differ. In \cite{wen2015performance}, a linear system is considered where the output is the result of the input multiplied by an unknown matrix, plus additive noise, while in \cite{wen2015joint,wen2016bayes} uniform quantization is added at the output.

%For a triple $(\txA,\rxA,\chA)$, where $\txv_\Tb\in\txA$ and $\rxv_\Tb\in\rxA$ are triangular processes connected through a distribution $p(\rxv_T|\txv_T;\chva{\Tb})$ parameterized by $\chva{\Tb}\in\chA$ independent of $\txv_{\Tb}$, the entropy $\EntR(\rxA|\txA_{\tau})$ is determined by $\cP_{\tau}$ defined in \eqref{eq:P_set}. 
As we now show, computations of $\EntR(\rxA_{\dy}|\txA)$ can be leveraged to compute $\EntR(\rxA_\dy|\txA_\dx)$ for $\dy,\dx>-1$. 
We consider the case when the input $\tx_t$ are \iid for all $t$, and the distribution set $\cP(\Tt,\tau)$ defined in \eqref{eq:P_set} can be simplified as
%and show that $\EntR(\rxA_{\dy}|\txA)$ can be used as a starting point to obtain $\EntR(\rxA_\dy|\txA_{\dx})$ for all $\dy,\dx>-1$
\begin{align}
    \cP(\Tt,\tau)=\{p(\rx|\tx;\chva{\frac{\Tt}{\tau}}),p(\chva{\frac{\Tt}{\tau}}),p(\tx)\}.
    \label{eq:P_set_iid_input}
\end{align}
%where $\tau$ is the fraction of the blocklength used for training, and the number of unknown parameters in the vector $\chva{\frac{\Tt}{\tau}}$ can depend on the blocklength $\frac{\Tt}{\tau}$.
%and stays constant within each blocklength. 
%The entropies $\EntR(\rxA_{\dy}|\txA)$ and $\EntR(\rxA_\dy|\txA_{\dx})$ are computed using the distribution $\cP(\Tt,\tau)$ in \eqref{eq:P_set_iid_input}, and they are functions of $\tau$.
The following theorem assumes that we have $\EntR(\rxA_\dy|\txA)$ available as a function of $(\tau,\dy)$ for all $\dy\geq 0$:

\begin{thm}
\label{thm:scale_joint_entropy}
Assume that A3 is met, $\tx_t$ are \iid for all $t$, $\EntR(\rxA_\dy|\txA)$ exists and is continuous in $\tau$ and $\dy$ for $\tau\in(0,1)$ and $\dy\in(-1,\frac{1}{\tau}-1]$. Define
\begin{align}
    F(\tau,\dy)=\EntR(\rxA_\dy|\txA),
    \label{eq:entropy_generator}
\end{align}
where $\dy\geq 0$ and $\EntR(\rxA_\dy|\txA)$ is defined in \eqref{eq:EntR_cond_def}.
Then
\begin{align*}
    \EntR(\rxA_\dy|\txA_{\dx})=
      (1+u)\cdot F\left((1+u)\tau,\frac{\dy-u}{1+\dx}\right), 
    \numberthis
    \label{eq:EntR_rx_gamma_tx_lambda}
\end{align*}
\comm{
\begin{align*}
    \EntR(\rxA_\dy|\txA_{\dx})=\begin{cases} 
      (1+\dy)\cdot F((1+\dy)\tau,0), &\dy\leq \dx; \\
      (1+\dx)\cdot F\left((1+\dx)\tau,\frac{\dy-\dx}{1+\dx}\right), & \dy> \dx.
      \end{cases}
    \numberthis
    \label{eq:EntR_rx_gamma_tx_lambda}
\end{align*}
}
for all $\dy,\dx\in(-1,\frac{1}{\tau}-1]$, where $u=\min(\dy,\dx)$.
\end{thm}
\begin{proof}
According to  \eqref{eq:EntR_cond_def} and \eqref{eq:entropy_generator}, we have
\begin{align*}
    \lim_{\Tt\to\infty}\frac{1}{\Tt}\Ent(\rxv_{\ceil{(1+\dy)\Tt}}|\txv_{\Tt})=F(\tau,\dy),
\end{align*}
%where the unknown parameter vector is $\chv_{\frac{\Tt}{\tau}}$.
which is computed using $\cP(\Tt,\tau)$ defined in \eqref{eq:P_set_iid_input}.  When $\dx\geq\dy>-1$, we have
\begin{align}
    &\EntR(\rxA_\dy|\txA_\dx)=\lim_{\Tt\to\infty}\frac{1}{\Tt}\Ent(\rxv_{\ceil{(1+\dy)\Tt}}|\txv_{\ceil{(1+\dx)\Tt}}) \label{eq:Hdef}\\
    =&\lim_{\Tt\to\infty}\frac{1}{\Tt}\Ent(\rxv_{\ceil{(1+\dy)\Tt}}|\txv_{\ceil{(1+\dy)\Tt}}). \nonumber
\end{align}
Let $\tilde{\Tt}=\ceil{(1+\dy)\Tt}$, and we have
\begin{align*}
    \EntR(\rxA_\dy|\txA_\dx)=\lim_{\Tt\to\infty}\frac{1+\dy}{\tilde{\Tt}}\Ent(\rxv_{\tilde{\Tt}}|\txv_{\tilde{\Tt}}).
    \numberthis
    \label{eq:change_of_variable}
\end{align*}
Since $\cP(\Tt,\tau)$ only depends on $\frac{\Tt}{\tau}$, we can rewrite $\cP(\Tt,\tau)$ as
\begin{align*}
    \cP(\Tt,\tau)=\cP(\tilde{\Tt},\frac{\tilde{\Tt}}{\Tt}\tau),
\end{align*}
and therefore \eqref{eq:entropy_generator} and \eqref{eq:change_of_variable} yield
\begin{align*}
   \EntR(\rxA_\dy|\txA_\dx) =& (1+\dy)\cdot F(\lim_{\Tt\to\infty}\frac{\tilde{\Tt}}{\Tt}\tau,0)\\
   =&(1+\dy)\cdot F((1+\dy)\tau,0).
   \numberthis
   \label{eq:expand_1}
\end{align*}

For $\dy>\dx> -1$,
%\begin{align*}
%    &\EntR(\rxA_\dy|\txA_\dx)=\lim_{\Tt\to\infty}\frac{1}{\Tt}\Ent(\rxv_{(1+\dy)\Tt}|\txv_{(1+\dx)\Tt}).
%\end{align*}
let $\tilde{\Tt}=\ceil{(1+\dx)\Tt}$, and then \eqref{eq:Hdef} yields
%and we can rewrite $\cP(\Tt,\tau)$ as
%\begin{align*}
%    \cP(\Tt,\tau)=\cP(\tilde{\Tt},(1+\dx)\tau).
%\end{align*}
%Then,
\begin{align*}
   &\EntR(\rxA_\dy|\txA_\dx) =\lim_{\tilde{\Tt}\to\infty}\frac{1+\dx}{\tilde{\Tt}}\Ent(\rxv_{\frac{\ceil{(1+\dy)\Tt}}{\ceil{(1+\dx)\Tt}}\tilde{\Tt}}|\txv_{\tilde{\Tt}})\\
    &=(1+\dx)\cdot F\left(\lim_{{\Tt}\to\infty} \frac{\tilde{\Tt}}{\Tt}\tau ,\lim_{{\Tt}\to\infty} \frac{\ceil{(1+\dy)\Tt}}{\ceil{(1+\dx)\Tt}}-1 \right)\\
    &=(1+\dx)\cdot F\left((1+\dx)\tau,\frac{\dy-\dx}{1+\dx}\right).
    \numberthis
   \label{eq:expand_2}
\end{align*}
By combining \eqref{eq:expand_1} and \eqref{eq:expand_2}, we obtain \eqref{eq:EntR_rx_gamma_tx_lambda}.
\end{proof}

The following example demonstrates how to apply the main theorems.
\subsection*{Example 5: Bit flipping through random channels}
\noindent
Let 
%$\tx_{\Tb,t}$ and $\rx_{\Tb,t}$ be the input and the output of a binary system with blocklength $\Tb$ :
\begin{equation}
    \rx_{t}=\tx_{t}\oplus \chs_{k_t},\quad t=1,\ldots,\frac{\Tt}{\tau},
    \label{eq:binary_XOR_channel}
\end{equation}
where the binary input $\tx_{t}$ is XOR'ed with a random bit $\chs_{k_t}$ intended to model the unknown ``state" of the channel $k_t$.  Thus, each channel either lets the input bit directly through, or inverts it.  The $\tx_{t}$ are \iid equally likely to be zero or one, Bernoulli($\frac{1}{2}$) random variables.  Let $a>0$ be a parameter, where $a\cdot\frac{\Tt}{\tau}$ is the (integer) number of possible unique channels whose states are stored in the vector $\chv_{\frac{\Tt}{\tau}}=[\chs_{1},\chs_{2},\cdots,\chs_{a\cdot\frac{\Tt}{\tau}}]^\Tp$ comprising \iid Bernoulli($\frac{1}{2}$) random variables that are independent of the input. The channel selections $\bk_{\Tt}=[k_1,\ldots,k_{\frac{\Tt}{\tau}}]$ are chosen as an \iid uniform sample from $\{1,2,\cdots,a\cdot\frac{\Tt}{\tau}\}$ (with repetition possible), and the choices are known to the transmitter and receiver.  We wish to send training signals through these channels to learn $\chv_{\frac{\Tt}{\tau}}$; the more entries of this vector that we learn, the more channels become useful for sending data, but the less time we have to send data before the blocklength $\frac{\Tt}{\tau}$ runs out and $\chv_{\frac{\Tt}{\tau}}$ changes.
We want to determine the optimum $\tau$ as $\Tt\to\infty$ using \eqref{eq:opt_training_tau}. We therefore compute $\EntR(\rxA_\dy|\txA)$ and then use Theorem \ref{thm:scale_joint_entropy} to obtain $\EntR(\rxA_\dy|\txA_{\dx})$,
which is used to compute $\cII{\txA}{\rxA}{}$ through Theorem \ref{thm:computation_of_MuI_from_derivative}. 

By definition, $\EntR(\rxA|\txA)=\lim\limits_{\Tt\to\infty}\frac{1}{\Tt}\Ent(\rxv_{\Tt}|\txv_{\Tt})$.
The model \eqref{eq:binary_XOR_channel} yields
\begin{align*}
    &\Ent(\rxv_{\Tt}|\txv_{\Tt}) \stackrel{(\rm a)}{=}\Ent(\{\chs_{k_1},\ldots,\chs_{k_\Tt}\}|\txv_{\Tt})\\
    \stackrel{(\rm b)}{=}&\Ent(\{\chs_{k_1},\ldots,\chs_{k_\Tt})
    \stackrel{(\rm c)}{=}\E_{\bk_{\Tt}}|A_\Tt|,
\end{align*}
where $A_{\Tt} = \{k_1,\ldots,k_\Tt\}$, $|\cdot|$ denotes the cardinality of a set, $^{(a)}$ uses $\chs_{k_t}=\rx_t\oplus\tx_t$, $^{(b)}$ uses the independence between $\txv_{\Tt}$ and $\chs_t$, $^{(c)}$ uses the independence between $\chs_t$ and $\chs_k$ when $t\neq k$.  Then $|A_{\Tt}|=\sum_{i=1}^{a\frac{\Tt}{\tau}} \mathbbm{1}_{(i\in A_{\Tt})}$, where $\mathbbm{1}_{(\cdot)}$ is the indicator function and
\begin{align*}
    &\Ent(\rxv_{\Tt}|\txv_{\Tt})=\E|A_{\Tt}|=\sum_{i=1}^{a\frac{\Tt}{\tau}}\E(\mathbbm{1}_{(i\in A_{\Tt})})\\
    &=\sum_{i=1}^{a\frac{\Tt}{\tau}} \Pr(i\in A_{\Tt})=\frac{a\Tt}{\tau}(1-\Pr(1\notin A_{\Tt}))\\
    &=\frac{a\Tt}{\tau}(1-\prod_{t=1}^{\Tt}\Pr(1\neq k_t))=\frac{a\Tt}{\tau}(1-(1-\frac{\tau}{a\Tt})^{\Tt}).
\end{align*}
Therefore,
\begin{align*}
    &\EntR(\rxA|\txA)=\lim_{\Tt\to\infty}\frac{a}{\tau}(1-(1-\frac{\tau}{a\Tt})^{\Tt})=\frac{a}{\tau}(1-e^{-\frac{\tau}{a}}).
    \numberthis
    \label{eq:XOR_tau_1}
\end{align*}
By the chain rule for entropy, we have
\begin{align*}
    \EntR&(\rxA_\dy|\txA)=\EntR(\rxA|\txA) \\
    &+ \lim_{\Tt\to\infty}\frac{1}{\Tt}\Ent(\rx_{\Tt+1},\rx_{\Tt+2},\ldots,\rx_{\ceil{(1+\dy)\Tt}}|\txv_{\Tt},\rxv_{\Tt}).
\end{align*}
Since
\begin{align*}
   \ceil{\dy\Tt}\geq&\Ent(\rx_{\Tt+1},\rx_{\Tt+2},\ldots,\rx_{\ceil{(1+\dy)\Tt}}|\txv_{\Tt},\rxv_{\Tt})\\
    \geq&\Ent(\rx_{\Tt+1},\rx_{\Tt+2},\ldots,\rx_{\ceil{(1+\dy)\Tt}}|\txv_{\Tt},\rxv_{\Tt},\chva{\frac{\Tt}{\tau}})\\
    =&\Ent(\tx_{\Tt+1},\tx_{\Tt+2},\ldots,\tx_{\ceil{(1+\dy)\Tt}}|\txv_{\Tt},\rxv_{\Tt},\chva{\frac{\Tt}{\tau}})\\
    =&\Ent(\tx_{\Tt+1},\tx_{\Tt+2},\ldots,\tx_{\ceil{(1+\dy)\Tt}})=\ceil{\dy\Tt},
\end{align*}
we conclude that 
\begin{align*}
    F(\tau,\dy)=\EntR&(\rxA_\dy|\txA)=\frac{a}{\tau}(1-e^{-\frac{\tau}{a}})+\dy.
    \numberthis
    \label{eq:tau_XOR_EntR}
\end{align*}
%For $\tau\in(0,1]$ and $\dy>0$, \eqref{eq:XOR_tau_1} and \eqref{eq:tau_XOR_EntR} yield
%Therefore
%\begin{align*}
%    F(\tau,\dy)=\frac{a}{\tau}(1-e^{-\frac{\tau}{a}})+\dy.
%    \numberthis
%    \label{eq:F_expression_XOR}
%\end{align*}
%where $F(\tau,\dy)$ is defined as in \eqref{eq:entropy_generator}.  
From \eqref{eq:binary_XOR_channel}, we have
\begin{align*}
    p(\rxv_{\frac{\Tt}{\tau}}|\txv_{\frac{\Tt}{\tau}};\chva{\frac{\Tt}{\tau}}) =\prod_{t=1}^{\frac{\Tt}{\tau}}p(\rx_t|\tx_t;\chva{\frac{\Tt}{\tau}}),
\end{align*}
where $p(\rx_t|\tx_t;\chva{\frac{\Tt}{\tau}})=\mathbbm{1}_{(\rx_t=\tx_{t}\oplus \chs_{k_t})}$ for all $t$. It is clear that Assumption A3 is met and $\tx_t$ are \iid independent of $\chv_{\frac{\Tt}{\tau}}$. Theorem \ref{thm:scale_joint_entropy} yields
\begin{align*}
    \EntR(\rxA_\dy|\txA_\dx) =  \begin{cases} 
      \frac{a}{\tau}(1-e^{-\frac{\tau}{a}(1+\dy)}), & \dy\leq\dx; \\
      \frac{a}{\tau}(1-e^{-\frac{\tau}{a}(1+\dx)})+(\dy-\dx), & \dx<\dy,
      \end{cases}
\end{align*}
for $\dy,\dx\in(-1,\frac{1}{\tau}-1)$.
Then, Corollary \ref{cor:A3_derivative_relation} yields
\begin{align*}
    \EntInnC{\txA}{\rxA}{\dy^+}{\dy} =\frac{\partial\EntR(\rxA_\dy|\txA_\dy)}{\partial\dy}=e^{-\frac{\tau}{a}(1+\dy)},
\end{align*}
\begin{align*}
    \EntInnC{\txA}{\rxA}{\dx}{\dy} =\frac{\partial\EntR(\rxA_\dy|\txA_\dx)}{\partial\dy}=  \begin{cases}
      e^{-\frac{\tau}{a}(1+\dy)}, & \dy<\dx; \\
      1, & \dy>\dx.
      \end{cases}
\end{align*}
Therefore, Assumption A1 holds, and Lemma \ref{lem:bound_for_A2} allows us to conclude that A2 also holds.

From Theorems \ref{thm:computation_of_MuI_from_derivative} and \ref{thm:MI_equal_gap}, we obtain
\begin{align*}
    \cII{\txA}{\rxA}{}
%    =& \lim_{\gamma\downto\tau}\frac{\partial\EntR(\rxA_\gamma|\txA_{\tau})}{\partial\gamma}-\lim_{\gamma\downto\tau}\frac{\partial\EntR(\rxA_\gamma|\txA)}{\partial\gamma}\\
    =&\;1-e^{-\frac{\tau}{a}}, \\
    \numberthis
    \label{eq:MuI_example_exp}
    \cIRA{\txA}{\rxA}{}\geq &\;(1-\tau)(1-e^{-\frac{\tau}{a}}).
\end{align*}
Finally, \eqref{eq:opt_training_tau} yields $\tauopt=\argmax\limits_{\tau}(1-\tau)(1-e^{-\frac{\tau}{a}})$, or
\begin{equation}
    \tauopt=\begin{cases} 
      -a\ln a, & a\to 0; \\
      \frac{1}{2}, & a\to\infty;\\
      \frac{1}{e}, & a=\frac{1}{e}.
 %     0.443, & a=1.
      \end{cases}
\end{equation}
When $a$ is small, $\tauopt$ is larger than $a$; when $a$ is large, $\tauopt$ saturates at $\frac{1}{2}$; and $a=\frac{1}{e}$ is the dividing line between $\tauopt>a$ and $\tauopt<a$.  The corresponding rates per receive symbol are
\begin{equation}
    \Ropt=\begin{cases} 
      (1+a\ln a)(1-a), & a\to 0; \\
      \frac{1}{2}(1-e^{-\frac{1}{2a}}), & a\to\infty;\\
      (1-\frac{1}{e})^2, & a=\frac{1}{e}.
 %     0.443, & a=1.
      \end{cases}
\end{equation}
These results indicate that the optimum fraction of the blocklength that should be devoted to training varies as a function of the number of possible unique channels.  When $a=1$, the number of unique channels equals the blocklength $B$, and $\tau$ that maximizes  \eqref{eq:MuI_example_exp} is approximately $0.44$.  For a large number of unique channels relative to the blocklength ($a\tendsto\infty$), the fraction of the training time saturates at $1/2$.  When $a$ is small, the optimum fraction of the blocklength devoted to training decreases to zero, but more slowly than $a$.

The next section generalizes our results from scalar inputs and outputs to higher dimensional objects and develops a channel coding theorem to provide operational significance to $(1-\tau)\cII{\txA}{\rxA}{}$.

\section{Generalization to High-dimensional Processes, and Channel Coding Theorem}
\label{sec:high_dimen_generalize}
\subsection{High-dimensional processes}
We define the input and output processes as $\txA=(\txv_1,\txv_2,\ldots)$ and $\rxA=(\rxv_{1},\rxv_{2},\ldots)$, which now comprise vectors, matrices, or tensors. For simplicity, we consider $\txv_t$ and $\rxv_{t}$ as vectors. 
%$\Tt$ known input-output pairs are used for training to learn the system unknown parameter set $\chA$, which is assumed constant during a blocklength $\frac{\Tt}{\tau}$ and then changed independently into another block. 
Denote $\txm_t=[\txv_1,\txv_2,\ldots,\txv_{t}]$ and $\txm^t=[\txv_{t},\txv_{t+1},\ldots,\txv_{\frac{\Tt}{\tau}}]$, and similarly for $\rxm_t$ and $\rxm^t$. The notation here differs from previous sections; we use $\txv_t$ as the $t$th vector in the process, and $\txm_t$ as the first $t$ vectors, and $\txm^{t+1}$ as the last $\frac{\Tt}{\tau}-t$ vectors.  The vectors $\rxv_t$ have length $N$, which can be a function of the blocklength $\frac{\Tt}{\tau}$.

Similarly to \eqref{eq:MuI_def}, \eqref{eq:EntInn_cond_def}, \eqref{eq:EntInn_cond_def_one_extra},  and \eqref{eq:EntR_cond_def},  we define
\begin{align*}
\EntInnC{\txA}{\rxA}{\dx}{\dy} =\lim_{\Tt\to\infty}&\frac{1}{\rn}\Ent(\rxv_{\ceil{(1+\dy)\Tt}+1}\\
&|\txm_{\ceil{(1+\dx)\Tt}},\rxm_{\ceil{(1+\dy)\Tt}}),
    \numberthis
    \label{eq:EntInn_cond_def_high}
\end{align*}
\begin{align*}
\EntInnC{\txA}{\rxA}{\dx^+}{\dy}
=\lim_{\Tt\to\infty}&\frac{1}{\rn}\Ent(\rxv_{\ceil{(1+\dy)\Tt}+1}\\
&|\txm_{\ceil{(1+\dx)\Tt}+1},\rxm_{\ceil{(1+\dy)\Tt}}),
    \numberthis
    \label{eq:EntInn_cond_def_one_extra_high}
\end{align*}
\begin{align}
    \cIInn{\txA}{\rxA}{}=\lim_{\Tt\to\infty}\frac{1}{\rn}\MuI(\txv_{\Tt+1};\rxv_{\Tt+1}|\txm_{\Tt},\rxm_{\Tt}),
    \label{eq:MuI_def_high}
\end{align}
\begin{equation}
    \EntR(\rxA_\dy|\txA_\dx)=\lim_{\Tt\to\infty}\frac{1}{\rn\Tt}\Ent(\rxm_{\ceil{(1+\dy)\Tt}}|\txm_{\ceil{(1+\dx)\Tt}}).
    \label{eq:EntR_cond_def_high}
\end{equation}
Similarly to \eqref{eq:cI_def_with_P_tau} and \eqref{eq:joint_MuI_thm}, we define
\begin{align*}
  \cII{\txA_\dy}{\rxA_\dy}{}= &\lim_{\Tt\to\infty}\frac{1}{\rn}\MuI(\txv_{\ceil{(1+\dy)\Tt}+1};\\
  &\rxv_{\ceil{(1+\dy)\Tt}+1}|\txm_{\ceil{(1+\dy)\Tt}},\rxm_{\ceil{(1+\dy)\Tt}}),
  \numberthis
  \label{eq:cI_def_with_P_tau_high}
\end{align*}
\begin{align*}
    \cIRA{\txA_\dy}{\rxA_\dy}{}=&\lim_{\Tt\to\infty}\frac{\tau}{\rn\Tt}\MuI(\txm^{\ceil{(1+\dy)\Tt}+1};\\
    &\rxm^{\ceil{(1+\dy)\Tt}+1}|\txm_{\ceil{(1+\dy)\Tt}},\rxm_{\ceil{(1+\dy)\Tt}}),
    \numberthis
    \label{eq:joint_MuI_thm_high}
\end{align*}
where $\cP(\Tt,\tau)$ is defined as in \eqref{eq:P_set}:
\begin{equation}
\cP(\Tt,\tau)=\{p(\rxv|\txv;\chma{\frac{\Tt}{\tau}}),p(\chma{\frac{\Tt}{\tau}}),p(\txm_\Tt),p(\txv_{\Tt+1})\},
\label{eq:P_set_high}
\end{equation}
and $\chma{\frac{\Tt}{\tau}}$ are the unknown parameters.

Theorems \ref{thm:derivative_monotonic}--\ref{thm:scale_joint_entropy}, Corollary \ref{cor:A3_derivative_relation}, and Lemma \ref{lem:bound_for_A2} can all be generalized.  We show only the generalization of Theorem \ref{thm:derivative_monotonic}.
\begin{thmbis}{thm:derivative_monotonic}
\label{thm_derivative_monotonic_high_dimension}
Let both $\EntR(\rxA_\dy)$ and its derivative with respect to $\dy$ exist.  Suppose there exists a $\kappa>0$ so that $\Ent(\rxv_{t+1}|\rxm_{t})$ is monotonic in $t$ when $t\in[\floor{(1+\dy-\kappa)\Tt},\ceil{(1+\dy+\kappa)\Tt}]$ as $\Tt\to\infty$.  Then
\begin{equation}
    \EntInn(\rxA_\dy) =\frac{\partial\EntR(\rxA_\dy)}{\partial\dy}.
    \numberthis
    \label{eq:derivative_2_high}
\end{equation}
\end{thmbis}

For the generalizations of Theorems \ref{thm:integral_thm_monotonic}--\ref{thm:scale_joint_entropy}, Corollary \ref{cor:A3_derivative_relation}, and Lemma \ref{lem:bound_for_A2}, the definitions in \eqref{eq:EntInn_cond_def_high}--\eqref{eq:joint_MuI_thm_high} are used.  
%and the conditional mutual information  \eqref{eq:MuI_estimate_lower_bound} are normalized by $N$.

\subsection{Channel coding theorem}
\label{subsec:channel_coding_thm_SISO}
We now provide an operational description of the mutual information inequality \eqref{eq:MuI_begin_lower_bound}. 
%We consider a communication system and develop a channel coding theorem that shows the limit of the rate that can be achieved for reliable communication. For simplicity, we consider systems with scalar input and output.
We consider a communication system where the channel is constant for blocklength $\frac{T}{\tau}$, and then changes independently and stays constant for another blocklength, and so on. The first $\Tt$ symbols of each block are used for training with known input and output.
Under Assumption A3, the communication system is memoryless, is time-invariant within each block, and the input is \iid independent of $\txv_{\Tt}$ after training.  The system is retrained with every block, and the message to be transmitted is encoded over the data phase of multiple blocks. 

A $(2^{nR\frac{\Tt}{\tau} },n,\Tt)$-code for a block-constant channel with blocklength $\frac{\Tt}{\tau}$ is defined as an encoder that maps a message $S\in\{1,2,\ldots,2^{nR\frac{\Tt}{\tau}}\}$ to the input in the data phase $\txv^{\Tt+1}$ among $n$ blocks,
and a decoder that maps $\txv_{\Tt}$, and the entire output $\rxv_{\frac{\Tt}{\tau}}$ for $n$ blocks to $\hat{S}\in\{1,2,\ldots,2^{nR\frac{\Tt}{\tau}}\}$. The code rate $R$ has units ``bits per transmission", and the maximum probability of error of the code is defined as
\begin{equation}
    \Pe(n,\Tt)=\max_{S}\Pr(\hat{S}\neq S).
    \label{eq:max_prob_err}
\end{equation}
The channel coding theorem is shown below.
\begin{thm}
\label{thm:channel_coding_thm_inf_Tb}
Assume A3 is met, with a channel that is constant with blocklength $\frac{\Tt}{\tau}$, and whose conditional distribution is parameterized by $\chva{\frac{\Tt}{\tau}}$ and is independent of the input.  If $\cII{\txA}{\rxA}{}$ exists, then for every $R$ that satisfies
\begin{equation}
    R<(1-\tau)\cII{\txA}{\rxA}{},
    \label{eq:rate_lower_bound}
\end{equation}
there exists $T_0>0$, so that for all $T>T_0$, we can find a code $(2^{nR\frac{\Tt}{\tau} },n,\Tt)$ with maximum probability of error $\Pe(n,\Tt)\to 0$ as $n\to\infty$.
\end{thm}
\begin{proof}
Define 
\begin{equation}
    \cR_{\Tt}=\frac{\tau}{\Tt}\MuI(\txv^{\Tt+1};\rxv_{\frac{\Tt}{\tau}}|\txv_{\Tt}).
\end{equation}
For any finite $\Tt$, according to the classical channel coding theorem \cite{cover2012elements,yeung2008information,effros2010generalizing}, for every $R<\cR_{\Tt}$, there exists a code $(2^{n R\frac{\Tt}{\tau}},n,\Tt)$ with maximum probability of error $\Pe(n,\Tt)\to 0$ as $n\tendsto\infty$. 

It is clear that $\txv^{\Tt+1}$ is independent of $(\txv_{\Tt},\rxv_{\Tt})$. Therefore, we have
\begin{equation}
    \cR_{\Tt}=\frac{\tau}{\Tt}\MuI(\txv^{\Tt+1};\rxv^{\Tt+1}|\txv_{\Tt},\rxv_{\Tt}).
\end{equation}

Since $\tx_{\Tt+1},\tx_{\Tt+2},\ldots,\tx_{\frac{\Tt}{\tau}}$ are \iid, and $p(\rx_t|\tx_t;\chva{\frac{\Tt}{\tau}})$ is a fixed conditional distribution for all $t=1,2,\ldots,\frac{\Tt}{\tau}$, we have \eqref{eq:monotonic_MuI_T} and \eqref{eq:chain_rule_MuI_bound}, which yield
\begin{equation*}
    \cR_{\Tt}\geq  (1-\tau)\MuI(\tx_{\Tt+1};\rx_{\Tt+1}|\txv_{\Tt},\rxv_{\Tt}).
    \numberthis
    \label{eq:bound_of_R}
\end{equation*}
According to the definition in \eqref{eq:cI_def_with_P_tau}, we have
\begin{align*}
     \cII{\txA}{\rxA}{}=\lim_{\Tt\to\infty}\MuI(\tx_{\Tt+1};\rx_{\Tt+1}|\txv_{\Tt},\rxv_{\Tt}).
\end{align*}
Therefore, for any $\kappa>0$, there exists a number $T_0>0$ so that when $\Tt>T_0$, we have
\begin{align*}
    \MuI(\tx_{\Tt+1};\rx_{\Tt+1}|\txv_{\Tt},\rxv_{\Tt})>\cII{\txA}{\rxA}{} - \kappa,
\end{align*}
and \eqref{eq:bound_of_R} yields
\begin{align*}
    \cR_{\Tt}>(1-\tau)(\cII{\txA}{\rxA}{} - \kappa),
\end{align*}
which means any rate $R\leq(1-\tau)(\cII{\txA}{\rxA}{} - \kappa)$ is achievable. 

By taking the limit $\kappa\downto 0$, we finish the proof.
\end{proof} 

This theorem shows that rates below $(1-\tau)\cII{\txA}{\rxA}{}$ are achievable when $\Tt$ is chosen large enough.  Only an achievability statement is given here since $(1-\tau)\cII{\txA}{\rxA}{}$ is a lower bound on $\cR_{\Tt}$ for large $\Tt$. 
%In general, both an achievability and converse apply to $\cR_{\Tb}(\tau)$ for large $\Tb$, which converges to $\cIR{\txA}{\rxA}{\tau}$ defined in \eqref{eq:cIR_def} in the limit, but we omit this discussion.

\section{Discussion and Conclusion}
\label{sec:discussion}

\subsection{Number of unknowns and bilinear model}
In general, a finite number of unknowns in the model leads to uninteresting results as $\Tt\tendsto\infty$.  For example, consider a system modeled as
\begin{equation}
    \rx_{t} = \chs\tx_t +v_t,\quad t=1,2,\ldots
    \label{eq:training_1D}
\end{equation}
where $\chs$ is the unknown gain of the system, $\tx_t$, $\rx_t$ are the input and corresponding output, $v_t$ is the additive noise, $\tau$ is the fraction of time used for training.  This system is bilinear in the gain and the input.  We assume that $\ns_t$ is modeled as \iid Gaussian $\cN(0,1)$, independent of the input. The training signals are $\tx_t=1$ for all $t=1,2,\ldots, \Tt$, and the data signals $\tx_t$ are modeled as \iid Gaussian $\cN(0,1)$ for all $t=\Tt+1,\Tt+2,\ldots$ An analysis similar to Example 5 produces
\begin{align*}
    \cIRA{\txA}{\rxA}{} \geq\frac{1-\tau}{2}\E_{\chs}\log(1+\chs^2),
\end{align*}
and therefore $\tauopt=0$ maximizes this bound. This result reflects the fact that $\chs$ is learned perfectly for any $\tau>0$ because there is only one unknown parameter for $\Tt$ training symbols as $\Tt\tendsto\infty$.  Hence, trivially, it is advantageous to make $\tau$ as small as possible.

More interesting is the ``large-scale" model
\begin{align*}
    \rxv_t=\bff(\chm\txv_t+\nv_t),\quad t=1,2,\ldots,
    \numberthis
    \label{eq:nonlinear_system_model}
\end{align*}
where $\txv_t$ and $\rxv_t$ are the $t$th input and output vectors with dimension $\tn$ and $\rn$, $\chm$ is an $\rn\times\tn$ unknown random matrix that is not a function of $t$, $\nv_1,\nv_2,\ldots$ are \iid unknown vectors with dimension $\rn$ and known distribution (not necessarily Gaussian), and $\bff(\cdot)$ applies a possibly nonlinear function $f(\cdot)$ to each element of its input. The training interval $\Tt$ is used to learn $\chm$.  Let $\tn$ and $\rn$ increase proportionally to the blocklength $\frac{\Tt}{\tau}$, and define the ratios
\begin{equation}
    \ratio=\frac{\rn}{\tn},\quad\beta=\frac{\Tt}{\tau\tn}.
    \label{eq:ratios_scale_up}
\end{equation}

This model can be used in large-scale wireless communication, signal processing, and machine learning applications. In wireless communication and signal processing \cite{takeuchi2010achievable,takeuchi2012large,hassibi2003much,takeuchi2013achievable,li2016much,li2017channel,wen2015performance,wen2015joint,wen2016bayes}, $\txv_t$ and $\rxv_t$ are the transmitted signal and the received signal at time $t$ in a multiple-input-multiple-output (MIMO) system with $\tn$ transmitters and $\rn$ receivers, $\chm$ models the channel coefficients between the transmitters and receivers, $\frac{\Tt}{\tau}$ is the coherence time during which the channel $\chm$ is constant, $\nv_t$ is the additive noise at time $t$, $f(\cdot)$ models receiver effects such as quantization in analog-to-digital converters (ADC's) and nonlinearities in amplifiers. A linear receiver, $f(x)=x$, is considered in \cite{takeuchi2010achievable,takeuchi2012large,takeuchi2013achievable,wen2015performance}. Single-bit ADC's with $f(x)=\sign(x)$ are considered in \cite{li2016much,li2017channel}, and low-resolution ADC's with $f(x)$ modeled as a uniform quantizer are considered in \cite{wen2015joint,wen2016bayes}.  The training and data signals can be chosen from different distributions, as in \cite{hassibi2003much,li2016much,li2017channel}.  Conversely, the training and data signals can both be \iid, as in \cite{takeuchi2013achievable,wen2015performance,wen2015joint,wen2016bayes}.
%and some information-theoretic lower bounds when $f(\cdot)$ is a single-bit quantizer include \cite{kang2019training,gao2019channel}.  [!!!Put this reference in Part II] 

%, $\chm$ is the wireless channel coefficients between the sources and the receive nodes, $\nv_t$ is the additive noise, and $f(\cdot)$ again models each receive node.
%Various receiver models are considered in \cite{shirazinia2015massive,shirazinia2016massive,serra2017distributed,choi2014channel,choi2015quantized}.  In \cite{shirazinia2015massive,serra2017distributed,choi2015quantized}, $\chm$ is assumed known perfectly at the fusion center. In general, $\chm$ needs to be estimated before data transmission, and we consider that the first $\Tt$ inputs are known signals for channel estimation, and the rest time $\Td$ are used for transmission of data generated at the sources. 

In machine learning, \eqref{eq:nonlinear_system_model} is a model of a single layer neural network (perceptron) \cite{opper1996statistical,engel2001statistical,shinzato2008learning} and $\txv_t$ is the input to the perceptron with dimension $\tn$, $\rxv_t$ is the scalar decision variable ($\rn=1$) at time $t$, $\chm$ holds the unknown weights of the perceptron, and $f(\cdot)$ is the nonlinear activation function. A perceptron is often used as a classifier, where the output of the perceptron is the class label of the corresponding input. In \cite{opper1996statistical,engel2001statistical}, \iid inputs are used to learn the weights, and orthogonal inputs are used in \cite{shinzato2008learning}. Binary class classifiers are considered in \cite{opper1996statistical,engel2001statistical,shinzato2008learning}.
Training employs $\Tt$ labeled input-output pairs $(\txv_t,\rxv_t)$, and the trained perceptron then classifies new inputs before it is retrained on a new dataset.  Generally, both the training and data are modeled as having the same distribution.

%In many cases, we may approximate a large-blocklength system by letting $\Tb\to\infty$ with $\tn,\rn,\Tt,\Td$ scaled proportionally as in \eqref{eq:ratios_scale_up}.  
To obtain optimal training results for \eqref{eq:nonlinear_system_model}, Theorems \ref{thm:computation_of_MuI_from_derivative}--\ref{thm:scale_joint_entropy} show that a starting point for computing
%$\EntR(\rxA_\gamma|\txA_{\tau}),\EntR(\rxA_\gamma|\txA),$
$\cII{\txA}{\rxA}{}$ is $\EntR(\rxA_\dy|\txA)$ for $\dy\geq 0$.  Fortunately, $\EntR(\rxA_\dy|\txA)$ results can sometimes be found in the existing literature; for example, in \cite{wen2015performance,wen2015joint,wen2016bayes}, $\EntR(\rxA_\dy|\txA)$ is used to calculate the mean-square error of the estimated input signal, conditioned on the training.  We may employ these same $\EntR(\rxA_\dy|\txA)$ results to quickly derive the training-based mutual information using the derivative analysis presented herein.  Part II of this paper focuses on this.
%We believe that $\EntR(\rxA|\txA_{\tau})$ calculations for other models can be leveraged using the framework provided herein to provide insight into training these mode

\subsection{Models for which assumptions are superfluous}

Assumptions A1 and A2 presented in the Introduction are likely superfluous for certain common system models, such as when the distributions on $\tx_t$ are \iid through the training and data phases, and the transition probabilities can be written as a product as in Assumption A3. However, as Lemma \ref{lem:bound_for_A2} shows, we have not yet characterized for which models A1 and A2 are automatically satisfied without additional assumptions on ${\cal H}'$, and think that this would be an interesting research topic for further work.

\comm{
We consider a common neural network design pattern in the area of machine learning, called {\it encoder-decoder architecture}\cite{cho2014properties,cho2014learning,badrinarayanan2017segnet,wang2017residual,wang2018reconstruction,lu2017knowing}, which is shown in Fig. \ref{fig:encoder_decoder_arch}. In such architecture, the input $\txA$ is encoded  into states through an encoder, which are often vectors or tensors, as the input to  a decoder, which outputs $\rxA$.

\begin{figure}
\includegraphics[width=3.2in]{encoder_decoder_arch.pdf}
\centering
    \caption{An encoder-decoder architecture in neural network design.}
    \label{fig:encoder_decoder_arch}
\end{figure}
}

\bibliographystyle{IEEEtran}
\bibliography{bib/refs.bib}

\end{document}